\documentclass[11pt]{article}
\usepackage{cite}
\usepackage{amsmath,amsfonts,amssymb}
\usepackage[small,bf,hang]{caption}
\usepackage{slashed}
\usepackage{mathabx}

\def\hybrid{
        \topmargin -20pt
        \oddsidemargin 0pt
        \headheight 0pt \headsep 0pt
        \textwidth 6.25in 
        \textheight 9.5in 
        \marginparwidth .875in
        \parskip 5pt plus 1pt \jot = 1.5ex}

\hybrid

 \csname
@addtoreset\endcsname{equation}{section}

\def\moth{\mathsurround=0pt}
\newdimen\zo \zo=0pt

\def\tick{\leaders\hrule height 0.5ex depth 0pt \hskip 0.5pt}
\def\upboxfill{$\moth \setbox\zo\hbox{\tick}%
  \hskip 3pt\hbox to 0pt{$\tick$\hss}\hrulefill \hbox to 7.5pt{$\tick$\hss}$}

\def\dtick{\leaders\hrule height .34pt depth 0.5ex \hskip 0.5pt}
\def\downboxfill{$\moth \setbox\zo\hbox{\dtick}%
  \hskip 2pt\hbox to 0pt{$\dtick$\hss}\hrulefill \hbox to 2pt{$\dtick$\hss}$}

\def\bec{\begin{center}}
\def\ec{\end{center}}

\def\cD{{\cal D}}

 \def\det{{\rm det\,}}
\def\be{\begin{equation}}
\def\ee{\end{equation}}
\def\bea{\begin{eqnarray}}
\def\eea{\end{eqnarray}}
\def\ba{\begin{array}}
\def\ea{\end{array}}

\def\ft#1#2{{\textstyle{{\scriptstyle #1}
\over {\scriptstyle #2}}}}

\begin{document}

\begin{titlepage}
\rightline{}
\rightline{\tt  LMU-ASC 45/13}
\rightline{July 2013}
\begin{center}
\vskip 1.0cm
{\LARGE \bf {U-duality covariant gravity}}\\
\vskip 2cm
{\large {Olaf Hohm${}^1$ and Henning Samtleben${}^2$}}
\vskip 1cm
{\it {${}^1$Arnold Sommerfeld Center for Theoretical Physics}}\\
{\it {Theresienstrasse 37}}\\
{\it {D-80333 Munich, Germany}}\\
olaf.hohm@physik.uni-muenchen.de
\vskip 0.2cm
{\it {${}^2$Universit\'e de Lyon, Laboratoire de Physique, UMR 5672, CNRS}}\\
{\it {\'Ecole Normale Sup\'erieure de Lyon}}\\
{\it {46, all\'ee d'Italie, F-69364 Lyon cedex 07, France}}\\
henning.samtleben@ens-lyon.fr

\vskip 1.5cm
{\bf Abstract}
\end{center}

\vskip 0.2cm

\noindent
\begin{narrower}
We extend the techniques of double field theory to more general gravity 
theories and U-duality symmetries, having in mind applications to the complete 
$D=11$ supergravity. In this paper we work out a $(3+3)$-dimensional `U-duality covariantization'  
of $D=4$ Einstein gravity, in which the Ehlers group $SL(2,\mathbb{R})$ 
is realized geometrically, acting in the $\bf{3}$ representation on half of the coordinates.   
We include the full $(2+1)$-dimensional metric, while  
the `internal vielbein' is a coset representative of $SL(2,\mathbb{R})/SO(2)$ 
and transforms under gauge transformations via generalized Lie derivatives. 
In addition, we 
introduce a gauge connection of the `C-bracket', 
and a gauge connection of $SL(2,\mathbb{R})$, albeit subject to constraints.      
The action 
takes the form of $(2+1)$-dimensional gravity coupled to a Chern-Simons-matter theory but  
encodes the complete $D=4$ Einstein gravity. We comment on  
generalizations, such as an `$E_{8(8)}$ covariantization' of M-theory.

\end{narrower}

\end{titlepage}

\newpage

\tableofcontents


\section{Introduction}
Duality symmetries play a distinguished role in string and M-theory. They are believed to 
be part of the `stringy gauge symmetry' that should govern the so far elusive fundamental 
formulation of string/M-theory. A better understanding of the geometrical nature of these duality 
symmetries may  give insights into the very geometry underlying string theory.   
The simplest  duality is T-duality that relates 
equivalent toroidal  string backgrounds $T^d$ via the non-compact group $O(d,d,\mathbb{Z})$
and also appears in the supergravity approximation as a continuous non-linearly realized  
global $O(d,d,\mathbb{R})$ symmetry. 
Double field theory is an approach to make this symmetry manifest at the level of the effective spacetime action  \cite{Siegel:1993th},
and our goal in this paper is to generalize the recent developments in  \cite{Hull:2009mi,Hull:2009zb,Hohm:2010jy,Hohm:2010pp}. 
(See also \cite{Tseytlin:1990va,Duff:1989tf,Siegel:1993xq} 
for earlier results.)  

Double field theory (DFT) introduces doubled coordinates transforming 
in the fundamental representation of $O(d,d)$ together with an $O(d,d)$ valued  `generalized metric'.
The extra coordinates are well-motivated from string theory, where they are dual to winding modes and, in fact, 
the cubic approximation to DFT has initially been derived from closed string field theory  \cite{Hull:2009mi,Kugo:1992md}. 
DFT provides, in particular, a strikingly simple formulation of the usual (super)gravity actions,   
including the heterotic theory \cite{Siegel:1993th,Hohm:2011ex}, 
massless and massive type II theories \cite{Hohm:2011zr,Hohm:2011dv,Hohm:2011cp,Jeon:2012kd}, and  
their supersymmetric extensions 
\cite{Siegel:1993th,Coimbra:2011nw,Hohm:2011nu,Jeon:2011sq,Jeon:2012hp}, and also leads to a compelling 
generalization of Riemannian geometry \cite{Siegel:1993th,Hohm:2010xe,Hohm:2011si,Hohm:2012gk,Hohm:2012mf,Jeon:2010rw,Jeon:2011cn}, 
which in turn is closely related to (and an extension of) results in the `generalized geometry' of  
Hitchin and Gualtieri \cite{Hitchin:2004ut,Gualtieri:2003dx,Gualtieri:2007bq}.  
(See \cite{Hohm:2011dz,Aldazabal:2011nj,Geissbuhler:2011mx,Grana:2012rr,Andriot:2012wx,Andriot:2012an,Aldazabal:2013mya} 
for other applications and 
\cite{Hohm:2011gs,Zwiebach:2011rg,Aldazabal:2013sca,Berman:2013eva} for reviews.) 

Given the close relation between 10-dimensional string theory and 11-dimensional M-theory  
it is natural to suspect that there should be similar extensions or reformulations of M-theory or, 
in its 2-derivative approximation, of $D=11$ supergravity, that renders U-duality symmetries 
manifest by introducing extra coordinates that transform under the U-duality group. 
Upon torus compactification, $D=11$ supergravity gives rise to exceptional 
symmetry groups such as  $E_{7(7)}$ in $D=4$ and $E_{8(8)}$ in $D=3$~\cite{Cremmer:1979up}. 
Already in the 1980's this spurred  interest in the question to what extent these structures are present in eleven dimensions. 
The work of de~Wit and Nicolai presents a reformulation of $D=11$ supergravity that abandons manifest 
11-dimensional covariance, using a Kaluza-Klein inspired $4+7$ or $3+8$ splitting of the 
coordinates, but which exhibits an enhanced local Lorentz symmetry in accordance with the 
(composite) gauge symmetries appearing in the $D=4$ or $D=3$ 
coset models \cite{deWit:1986mz,Nicolai:1986jk}. However, it did 
not manifest the exceptional groups, and further work in \cite{Koepsell:2000xg} suggested that 
additional coordinates should be introduced in order to achieve this, an idea that also features 
prominently in the proposal of \cite{West:2003fc}.  
Later work in \cite{Hillmann:2009ci} gave 
a manifestly $E_{7(7)}$ covariant formulation for a certain 7-dimensional truncation of $D=11$ 
supergravity by introducing coordinates in the ${\bf 56}$ of $E_{7(7)}$. 

The purpose of this paper is to show that it is possible  
to reformulate complete gravity theories in a U-duality covariant manner.  
We will follow a strategy similar to the one employed by de~Wit--Nicolai: 
we decompose the fields and coordinates  \`a la Kaluza-Klein \textit{without truncation} and then reorganize 
them, however, now in a way that is fully U-duality covariant by virtue of the extra coordinates.   
In addition, we will have to introduce extra fields and constraints, but the extra fields can be 
eliminated once the constraints are solved. 
After the advent of DFT, there have already been quite a number of papers extending the techniques 
developed here to various U-duality groups 
\cite{Berman:2010is,Berman:2011pe,Berman:2011cg,Coimbra:2011ky,Berman:2012vc,
Berman:2012uy,Cederwall:2013naa} (see also \cite{Hull:2007zu,Pacheco:2008ps} for earlier results). 
The actions given in this context exhibit manifest $E_{n(n)}$ symmetry for $n\leq 7$
and describe truncations of $D=11$ supergravity. More precisely, 
$D=11$ supergravity is truncated by setting to zero the off-diagonal components 
of the metric and of the 3-form, assuming that all fields depend 
only on `internal' coordinates,  and freezing the external metric to be the flat Minkowski metric (sometimes 
up to a warp factor).  
In terms of the more general gravity actions  to be introduced here this 
truncation amounts to eliminating all but one term, the `potential' term. However,  
the detailed relation of our results to those of 
\cite{Berman:2010is,Berman:2011pe,Berman:2011cg,Coimbra:2011ky,Berman:2012vc}
is not entirely transparent, as we briefly discuss below.

Trying to write a complete U-duality covariant gravity theory one encounters two (related) 
obstacles: 
\begin{itemize}
\item[(i)]  The off-diagonal field components (as the Kaluza-Klein vector originating from the metric)
do not naturally fit into the generalized metric that is used in DFT to write the action.  
\item[(ii)] In order to manifest the duality symmetries in lower dimensions it is typically 
necessary to dualize some of the off-diagonal field components into forms of lower rank. 
Such transformations are specific to a given dimension, and so it is not clear 
how to employ the required dual fields in, say, the full $D=11$ supergravity. 
\end{itemize}
For definiteness we consider in this paper a $3+n$ decomposition, which is 
appropriate for the case of $D=3$ duality symmetries. For $n=8$ the duality group is $E_{8(8)}$, 
the case relevant for 11-dimensional supergravity, while here we restrict ourselves to the simplest 
toy model, $n=1$, relevant for $D=4$ Einstein gravity, for which the duality group is the Ehlers group $SL(2,\mathbb{R})$. 
The $D=3$ case is particularly interesting for various reasons. In $D=3$ the Kaluza-Klein vector needs to be dualized into a scalar, 
which together with the Kaluza-Klein dilaton then parametrizes the $SL(2,\mathbb{R})/SO(2)$
coset space~\cite{Ehlers:1957}.  
Since the Kaluza-Klein vector originates from the metric, from a $D=4$ perspective 
this is like dualizing (part of) the graviton, something that due to the no-go results of  \cite{Bekaert:2002uh} is 
usually considered to be impossible. Indeed, previous papers  on the subject  
have unanimously concluded that, presumably for this reason, the 
$D=3$ case cannot be incorporated into a U-duality covariant framework 
\cite{Coimbra:2011ky,Berman:2012vc,Godazgar:2013rja}. 
However, it turns out that the 
techniques to deal with dual fields in gauged supergravity developed in 
\cite{Nicolai:2003bp,deWit:2003ja} are quite sufficient to address this problem, a fact that   
has already been employed a while ago in \cite{Hohm:2005sc,Hohm:2006ud}, which will be crucial for our construction. 
This resolution of the `dual graviton problem'  
(which can also be employed in a fully covariant framework \cite{West:2002jj,Boulanger:2008nd,Bergshoeff:2009zq}) 
may appear somewhat trivial, but as we will see is exactly 
what is needed in order to achieve a duality covariant formulation. 
While in this paper we will restrict ourselves to the $3+n$ decomposition, we expect that along similar lines,
using the techniques of gauged supergravity in generic dimensions, there will be formulations of 
the complete 11-dimensional supergravity that are covariant with respect to various U-duality groups. 

The $SL(2,\mathbb{R})$ covariant formulation of $D=4$ Einstein gravity to be developed in this paper 
introduces coordinates $Y^M$ in the ${\bf 3}$ of $SL(2,\mathbb{R})$, $M=1,2,3$, which is the adjoint representation 
or, equivalently, the fundamental representation of the isomorphic group $SO(1,2)$.\footnote{This choice is motivated by the observation that the gauge vectors, which naturally couple 
to the extended derivatives, typically live in the adjoint representation 
of the duality group in $D=3$ gauged supergravity.}
As in DFT we have to subject the theory to a (covariant) `section constraint' that effectively implies that among 
the three coordinates $Y^M$ only one is physical, which then completes the remaining $2+1$ coordinates 
to those of $D=4$ gravity. The constraints take the form 
 \be\label{sectioncon}
  \eta^{MN} \partial_M\otimes \partial_N \ = \ 0\;, \qquad 
  f^{MNK} \partial_N\otimes \partial_K \ = \ 0\;, \qquad 
  \ee 
where we introduced the Cartan-Killing form $\eta_{MN}$ of $SL(2,\mathbb{R})$ 
(or, equivalently, the $SO(1,2)$ invariant metric) and its structure constants $f^{MNK}$. 
Here, the notation $\otimes$ indicates that the differential operator annihilates 
all fields, but also all of their products. The first constraint in (\ref{sectioncon}) takes the same 
form as the `strong constraint' in DFT, but with the $O(d,d)$ metric replaced by the $SO(1,2)$
metric.  The second constraint has appeared in an analogous form in other 
U-duality covariant formulations \cite{Berman:2012vc}. Its addition in (\ref{sectioncon})
actually does not make the 
first constraint any stronger, for the first one implies already that all fields depend only on
one of the $Y^M$ coordinates, which then automatically solves the second constraint. 

As in DFT we introduce a generalized metric ${\cal M}_{MN}$ that, in a $D=3$ language, encodes the 
scalar fields. Alternatively, we can introduce a frame field ${\cal V}_{M}{}^{A}$, 
with inverse ${\cal V}_{A}{}^{M}$, 
subject to local $SO(2)$ transformations from the right, and define ${\cal M}={\cal V} {\cal V}^T$. 
These fields transform under gauge transformations with a parameter $\Lambda^M$ 
that is the $SL(2,\mathbb{R})$ covariant extension of the 4th diffeomorphism parameter. 
It acts on the fields via the generalized Lie derivative 
 \be\label{genLie}
  \delta_{\Lambda}{\cal V}_{A}{}^{M} \ = \ \big[ \Lambda,{\cal V}_{A}\big]_{D}^{M} \ \equiv \ 
  \Lambda^N\partial_N {\cal V}_{A}{}^{M}+\big(\partial^M\Lambda_N-\partial_N\Lambda^M\big){\cal V}_{A}{}^{N}\;, 
 \ee
where we introduced the analogue of the `D-bracket' in DFT (again with $O(d,d)$ replaced by $SO(1,2)$), 
which in turn reduces to the Dorfman bracket of 
generalized geometry when the dependence on the extra coordinates is dropped. 
The D-bracket is not antisymmetric. 
Its antisymmetrization is the C-bracket that governs the gauge algebra of generalized Lie derivatives, 
and which in the $O(d,d)$ case reduces to the Courant bracket of generalized geometry 
when there is no dependence on extra coordinates. It does not define a Lie algebra, because 
it does not satisfy the Jacobi identity; however, its `Jacobiator' is of a particular exact form.

In our formulation, all fields depend on the $Y^M$, but also on the `external' spacetime coordinates 
$x^{\mu}$, e.g., ${\cal V}={\cal V}(x,Y)$. The transformations (\ref{genLie}) are gauge transformations 
from the $(2+1)$-dimensional perspective in that the parameter $\Lambda^M$ depends on $x$. 
Therefore we also need to introduce a gauge vector $A_{\mu}{}^M$ that gauges (\ref{genLie}) and 
which is the $SL(2,\mathbb{R})$ covariant version of the Kaluza-Klein vector. It 
transforms as 
 \be
  \delta_{\Lambda}A_{\mu}{}^{M} \ = \ 
   \partial_{\mu}\Lambda^M+\big[\Lambda,A_{\mu}\big]_{D}^{M}\;. 
 \ee  
Formally, this is the usual Yang-Mills gauge transformation, but the bracket does 
not define a Lie algebra, so this is not a conventional gauge connection. 
This gauge field can still be used, however, to define covariant derivatives, so that, e.g., 
$D_{\mu}{\cal V}_{A}{}^{M}$ transforms covariantly under (\ref{genLie}). 
Due to the failure of the C-bracket to satisfy  the Jacobi identity, the naive field strength 
 \be
  F_{\mu\nu}{}^{M} \ = \ \partial_{\mu}A_{\nu}{}^{M}-\partial_{\nu}A_{\mu}{}^{M}-\big[A_{\mu},A_{\nu}\big]_{C}^{M}\;, 
 \ee
does not transform covariantly. However, its failure to transform covariantly is such that 
by the section constraint (\ref{sectioncon}) it is covariant when contracted with $\partial_M$, 
 \be
   \delta_{\Lambda}F_{\mu\nu}{}^{M}\otimes \partial_M \ = \ \big[\Lambda,F_{\mu\nu}\big]_{D}^M\otimes \partial_M \;.
 \ee

Due to the lack of covariance of $F_{\mu\nu}{}^{M}$ we cannot write an invariant action for 
$A_{\mu}{}^{M}$ alone. For this and other reasons it turns out to be necessary to introduce
a second gauge vector $B_{\mu M}$, which  can be viewed as a gauge connection for 
$SL(2,\mathbb{R})$. Naively this appears to introduce too much gauge symmetry because  
we would then seem to be able to gauge ${\cal M}_{MN}$ to the unit matrix. 
However, $B$ and its gauge parameter will actually have to satisfy some (covariant) constraints inherited from 
(\ref{sectioncon}), which effectively reduces the number of components of  $B_{\mu M}$
and the amount of gauge symmetry. We will discuss this in detail below. The additional constraints can be motivated 
from the observation that, on-shell and to lowest order, $B_{\mu M}$ 
is determined to be dual to a Noether current of 
the coset space sigma model, schematically 
$\star d B^M \sim  \partial^M{\cal M}^{-1}\,\partial {\cal M}$. Contracting this relation with 
$\partial_M$ it is only consistent with the section constraint (\ref{sectioncon}) if we 
also require $B_{\mu}{}^{M}\partial_M=0$. Given this constraint, we can now write 
a gauge invariant action, the Chern-Simons 3-form $B_M\wedge F^M$. 
This coupling is also needed in order to guarantee the on-shell equivalence with 
conventional Einstein gravity: after solving the section constraints  $B_{\mu M}$
becomes an auxiliary field whose field equation implies the duality relation between 
$F_{\mu\nu}$ and the dual scalar (being the only remnant of the `dual graviton').

The complete U-duality covariant gravity action is given by 
  \be\label{visionaryActionIntro}
  S \ = \ \int d^3x \, d^3Y \left(e\,\widehat{R}-\frac{1}{2\sqrt{2}}\,\varepsilon^{\mu\nu\rho}B_{\mu M}  F_{\nu\rho}{}^{M} 
  +\frac{1}{16}e\,g^{\mu\nu}
  {\cal D}_{\mu}{\cal M}^{MN}{\cal D}_{\nu}{\cal M}_{MN} -e\,V({\cal M},g) \right)\;, 
 \ee 
c.f.~(\ref{visionaryAction}) below. 
Here, all fields depend on the $D=3$ spacetime coordinates $x^{\mu}$ and the $Y^M$. 
The first term is the usual $D=3$ Einstein-Hilbert term, but with all partial derivatives replaced 
by covariant derivatives with respect to $A$ and an additional improvement of the Riemann tensor 
that is necessary in order to render the $D=3$ local Lorentz transformations a symmetry in 
presence of  $\partial_M$ derivatives. The potential $V$ reads 
 \be\label{scalarIntro}
 \begin{split}
  V({\cal M},g)& \ = \ -\frac{3}{16}\Big( {\cal M}^{KL}\partial_K {\cal M}^{MN}\partial_L {\cal M}_{MN}
  -4{\cal M}^{KL}\partial_K {\cal M}^{MN}\partial_N {\cal M}_{ML}\Big) \\
  &-\frac{1}{2} g^{-1}\partial_M g\, \partial_N{\cal M}^{MN}-\frac{1}{4} {\cal M}^{MN}g^{-1}\partial_Mg\,g^{-1}\partial_Ng
  -\frac{1}{4}  {\cal M}^{MN}\partial_M g^{\mu\nu}\,\partial_N g_{\mu\nu}\;. 
 \end{split}
 \ee 
The terms in the first line agree precisely with the corresponding terms in the DFT action, 
particularly the relative coefficient. The terms in the second line resemble the dilaton 
couplings in DFT, with $g=|\det g|$ playing the role of the dilaton. There is one novelty, 
however, in that the full $(2+1)$-dimensional metric $g_{\mu\nu}$ enters the last term. 
The action (\ref{visionaryActionIntro}) takes the form of $(2+1)$-dimensional gravity coupled to 
a Chern-Simons-matter theory. However, if we solve the section constraint 
by setting $\partial_M=(\partial_y,0,0)$, the action (\ref{visionaryActionIntro}) will be shown to be 
exactly equivalent to the $D=4$ Einstein-Hilbert action. 
All symmetries are manifest, except for the $(2+1)$-dimensional 
diffeomorphisms that are generated by a parameter $\xi^{\mu}(x,Y)$ that depends also on 
$Y$. In fact, it is this symmetry that uniquely fixes all relative coefficients in (\ref{visionaryActionIntro}).

This paper is organized as follows.  In sec.~2 we introduce the required background material 
from DFT, including the generalized Lie derivative and the D- and C-bracket. 
Based on this we present a generalization of Yang-Mills theory, with gauge connections 
based on the D- and C-bracket algebra rather than a Lie algebra, leading to a structure that 
resembles the tensor hierarchy in gauged supergravity. Then we introduce the $SL(2,\mathbb{R})$
gauge field $B_{\mu M}$ and discuss its constraints. In sec.~3 we define the $(3+3)$-dimensional theory, 
systematically introducing the Chern-Simons term, the scalar kinetic term and potential and the 
covariantized Einstein-Hilbert term. In sec.~4 we discuss the $(2+1)$-dimensional 
diffeomorphisms parametrized by $\xi^{\mu}(x,Y)$, which tie together the various terms. 
Finally, in sec.~5 we prove that upon 
solving the section constraint the theory is precisely equivalent to $D=4$ Einstein gravity. 
We conclude with an outlook in sec.~6, discussing possible generalizations such as 
to the $E_{8(8)}$ covariant formulation of 11-dimensional supergravity.

\section{Algebraic structures}

\subsection{Generalities}
We start by recalling some central concepts inspired by DFT. Instead of the 
T-duality group we consider the group $SL(2,\mathbb{R})\cong SO(1,2)$, whose invariant 
Cartan-Killing form we choose to be of signature $(- + +)$, 
 \be
  \eta_{MN} \ = \ \begin{pmatrix}    0 & 0 & 1 \\[0.5ex]
  0 & 1 & 0 \\[0.5ex]  1 & 0 & 0 \end{pmatrix}\;, 
  \label{eta}
 \ee 
where $M,N=1,2,3$ label the ${\bf 3}$ representation. The structure constants of $SL(2,\mathbb{R})$
can be written in terms of the Levi-Civita symbol,  
 \be\label{STRk}
  f_{MNK} \ = \ \varepsilon_{MNK} \;, 
 \ee 
which implies standard identities like $f^{MKL} f_{MPQ} = -2\delta^{[K}_{\;\;P}\,\delta^{L]}_{\;\;Q}$.

We introduce coordinates $Y^M$ in the ${\bf 3}$ representation, 
with dual derivatives $\partial_M$. As in DFT, the theory is subject to the `strong constraint'
 \be\label{weakerstrong}
  \eta^{MN}\partial_M \partial_N A\ = \ 0\;, \qquad \eta^{MN}\partial_M A\,\partial_N B \ = \ 0\;,  
 \ee
for arbitrary $A$, $B$.  In fact, with $\partial_M$ in the adjoint representation of $SL(2,\mathbb{R})$,
this constraint turns out to imply another seemingly stronger constraint
\be\label{stronger}
f^{KMN}\partial_M A\,\partial_N B \ = \ 0\;,
\ee
with the antisymmetric structure constants of the $SL(2,\mathbb{R})$ algebra. 
It will sometimes be convenient
to encode (\ref{weakerstrong}) and (\ref{stronger}) into a single equation of the type
\bea
\mathbb{P}_{KL}{}^{MN}\,\partial_M \otimes \partial_N  \ = \ 0\;,
\eea
with a projector of the form
\bea
\mathbb{P}_{KL}{}^{MN} &\equiv& \frac13\,\eta_{KL}\eta^{MN} -\frac12\, f_{KLP}f^{MNP}
\;.
\label{projector}
\eea
 
Next we introduce the generalized Lie derivative $\widehat{\cal L}_{\Lambda}$ that governs gauge transformations with 
respect to a vector parameter $\Lambda^M$. On a vector $V^M$ it reads 
 \be\label{GennnLie}
  \delta_{\Lambda}V^M \ = \ 
  \widehat{\cal L}_{\Lambda}V^M  \ \equiv \ \Lambda^N\partial_N V^M +\big(\partial^M\Lambda_N-\partial_N\Lambda^M\big)V^N\;, 
 \ee  
where here and in the following all indices are raised and lowered with $\eta_{MN}$. The terms 
on the right-hand side are also denoted as the `D-bracket' so that we also write 
 \be
  \delta_{\Lambda}V^M \ = \ \big[\Lambda,V\big]_{ D}^M\;.
 \ee
The generalized Lie derivative acts similarly on higher tensors, with each index rotated as 
in the  second term in  (\ref{GennnLie}). 
We note that due to the constraint (\ref{weakerstrong}), parameters of the form $\Lambda^M=\partial^M\chi$
do not generate gauge transformations, and we will refer to such gauge parameters as `trivial'. 
 
The gauge transformations governed by generalized Lie derivatives (\ref{GennnLie}) close according 
to the `C-bracket', 
 \be\label{CBRacket}
  \big[ \widehat{\cal L}_{\Lambda_1}, \widehat{\cal L}_{\Lambda_2}\big] \ = \  \widehat{\cal L}_{[\Lambda_{1},\Lambda_2]_{ C}}\;, 
 \ee
where 
 \be
  \big[\Lambda_1,\Lambda_2\big]_{ C}^{M} \ = \ \Lambda_1^N\partial_N\Lambda_2^M-\frac{1}{2}\Lambda_{1N}
  \partial^M\Lambda_2^N-(1\leftrightarrow 2)\;.
 \ee
The C-bracket is the antisymmetrization of the D-bracket in that the D-bracket differs from  the 
antisymmetric C-bracket by a symmetric  term, 
 \be\label{SymmC}
  \big[V,W\big]^M_{ C} \ = \   \big[V,W\big]^M_{ D}-\frac{1}{2}\partial^M\big(V^N W_N\big)\;.
 \ee 
Crucially, the C-bracket does not satisfy the Jacobi identity. Rather, there is a non-trivial Jacobiator, 
 \be\label{Jacobiator}
  \big[\big[U,V\big]_{C},W\big]_{C}^{M}+{\rm cycl.} \ = \ \frac{1}{6}\partial^M\left(\big[U,V\big]_{C}^{N}W_{N}+ {\rm cycl.}\right)\;.
 \ee 
Note that, although non-zero, the Jacobiator is of a trivial form and therefore does not generate 
gauge transformations, in agreement with the fact that the symmetry variations $\delta_{\Lambda}$
of fields always satisfy the Jacobi identity.  
 
We now discuss various objects that are tensorial in the generalized sense of (\ref{GennnLie}).  
First, the scalar fields are encoded by an $SL(2,\mathbb{R})$ vector transforming according to (\ref{GennnLie}) 
under gauge transformations. More precisely, they are given by  
a coset representative ${\cal V}_{M}{}^{A}$ of $SL(2,\mathbb{R})/SO(2)$, which is subject to global and local transformations  
 \be
  {\cal V}(Y)\; \rightarrow \; {\cal V}'(Y') \ = \ 
  g^T \,{\cal V}(Y)\,h(Y)\;, \qquad h(Y)\in SO(2)\;, \quad g\in SO(1,2)\;, 
  \label{matrixV}
 \ee
where $Y'=gY$.  In the following we will mainly work with the generalized metric 
${\cal M}_{MN}=({\cal V}\,{\cal V}^T)_{MN}$,
so that all expressions are 
manifestly invariant under local $SO(2)$ transformations. 
As in DFT, we have a second metric, $\eta_{MN}$, of different signature. 
Since this metric is used in the generalized Lie derivative (\ref{GennnLie}) to raise 
and lower indices, it is easy to see that acting on $\eta_{MN}$ itself the generalized Lie derivative is zero, 
 \be
  \widehat{\cal L}_{\Lambda}\eta_{MN} \ = \ 0\;.
 \ee
In the $SL(2,\mathbb{R})$ invariant formulation to be developed here there is another invariant  
tensor, given by the structure constants (\ref{STRk}) or the epsilon symbol.
To see that this is indeed an invariant tensor under generalized Lie derivatives, we compute first   
 \be
  \widehat{\cal L}_{\Lambda}\varepsilon_{MNK} \ = \ \Lambda^P\partial_P \varepsilon_{MNK}
  +3\big(\partial_{[M}\Lambda^P-\partial^P\Lambda_{[M}\big)
  \varepsilon_{NK]P}\;. 
  \ee 
With the Schouten identity $  \partial_{[M}\Lambda^P \varepsilon_{PNK]} = 0 $
we have 
 \be\label{firstSch}
  \partial_M\Lambda^P\,\varepsilon_{PNK}  +\partial_N\Lambda^P\,\varepsilon_{MPK} +\partial_K\Lambda^P\,\varepsilon_{MNP} \ = \ \partial_P\Lambda^P
  \varepsilon_{MNK}\;, 
 \ee  
and similarly with $-\partial^P\Lambda_{[M}\varepsilon_{PNK]}=0$ we find 
 \be\label{secSch}
  -\partial^P\Lambda_{M}\,\varepsilon_{PNK} -\partial^P\Lambda_{N}\,\varepsilon_{MPK} -\partial^P\Lambda_{K}\,\varepsilon_{MNP} \ = \ -\partial_P\Lambda^P \, 
  \varepsilon_{MNK} \;. 
 \ee 
Thus, the terms in the generalized Lie derivative of $\varepsilon_{MNK}$ cancel and we conclude
 \be\label{epscov}
  \widehat{\cal L}_{\Lambda} \varepsilon^{MNK} \ = \ \widehat{\cal L}_{\Lambda} f^{MNK} \ = \ 0\;. 
 \ee 
Therefore, both the $SL(2,\mathbb{R})$ metric $\eta_{MN}$ and the structure constants $f^{MNK}$ are 
gauge invariant. Note that the cancellation between (\ref{firstSch}) and (\ref{secSch}) was due to the
antisymmetric combination of $\partial\Lambda$ entering the Lie derivative. In contrast, 
in conventional geometry there is no such cancellation, so that the epsilon tensor is a tensor 
density rather than a strictly invariant tensor.

\subsection{Covariant derivatives for the D- and C-bracket}
As explained in the introduction, in our formulation all fields depend not only on $Y^M$ but also 
the $(2+1)$-dimensional spacetime coordinates $x^{\mu}$. In particular, a gauge parameter such as $\Lambda^M$ depends 
on $x^{\mu}$, and so from the perspective of the external space the transformations (\ref{GennnLie}) 
are gauge transformations. A spacetime derivative such as $\partial_{\mu}{\cal V}$ then 
does not transform covariantly with the generalized Lie derivative and therefore we have to 
introduce a gauge connection $A_{\mu}{}^{M}$ and covariant derivatives, as we will do in 
this section. The structure is completely analogous to that in DFT, which we recently 
investigated in \cite{Hohm:2013new1}. Here we summarize the main results and refer to 
 \cite{Hohm:2013new1} for detailed derivations. 

We start with the gauge transformations of  $A_{\mu}{}^{M}$, which in analogy to ordinary 
Yang-Mills theory we define to be 
 \be\label{betterdelta}
  \delta_{\Lambda}A_{\mu}{}^{M} \equiv \ \partial_{\mu}\Lambda^M+\big[\Lambda,A_{\mu}\big]_{D}^{M}
  \ = \ \partial_{\mu}\Lambda^M-\big[A_{\mu},\Lambda\big]_{D}^{M}+\partial^M\big(\Lambda^N A_{\mu N})\;. 
 \ee
Since the D-bracket is not antisymmetric, we had to employ (\ref{SymmC}) in order 
to reverse the arguments. We see that the two `natural' ways to write the gauge transformations 
\`a la Yang-Mills differ by a total $\partial^M$ derivative. As we will explain below, this difference is 
irrelevant due to an extra shift gauge symmetry on $A_{\mu}{}^{M}$.  
Similarly, we could have also written the transformation with the C-bracket. 
Explicitly, the gauge transformations can be written as 
 \be\label{covder}
   \delta_{\Lambda}A_{\mu}{}^{M} \ = \ \partial_{\mu}\Lambda^M+\Lambda^N \partial_N A_{\mu}{}^{M}
  +\left(\partial^M \Lambda_N-  \partial_N \Lambda^{M}\right)A_{\mu}{}^{N} \;, 
  \ee  
which shows that this is the covariant transformation plus the inhomogeneous term $\partial_{\mu}\Lambda$.   
With the gauge field $A_{\mu}{}^{M}$ we can next define a covariant $x^{\mu}$-derivative, 
which reads 
 \be\label{covDerGen}
  D_{\mu} \ = \ \partial_{\mu}-\widehat{\cal L}_{A_{\mu}}\;.  
 \ee 
Here, the generalized Lie derivative acts in the representation of the object on which $D_{\mu}$
acts. Despite the slightly non-standard form of the gauge transformations of the gauge fields, 
these derivatives are fully covariant under local $\Lambda^M$ transformations. 
Let us finally specialize (\ref{covDerGen}) to the 
covariant derivative for the scalars encoded by ${\cal M}_{MN}$, which reads explicitly 
   \be\label{covM}
    D_{\mu}{\cal M}_{MN} \ = \ \partial_{\mu}{\cal M}_{MN}-A_{\mu}{}^{K}\partial_K {\cal M}_{MN}
    -2\left(\partial_{(M} A_{\mu}{}^{K}-\partial^KA_{\mu (M}\right){\cal M}_{N)K}  \;.
   \ee 

We now turn to the field strength of $A_{\mu}{}^{M}$, which like in Yang-Mills theory we define as 
 \be\label{fieldstr}
  F_{\mu\nu}{}^{M} \ = \ \partial_{\mu}A_{\nu}{}^{M}-\partial_{\nu}A_{\mu}{}^{M}-\big[A_{\mu},A_{\nu}\big]_{C}^{M}\;. 
 \ee
As usual, the field strength emerges through the commutator of covariant derivatives,  
 \be\label{commF}
  \big[D_{\mu},D_{\nu}\big] \ = \ -\widehat{\cal L}_{F_{\mu\nu}}\;.
 \ee 
Since the C-bracket does not satisfy the Jacobi identity, $F_{\mu\nu}{}^{M}$ does not transform fully covariantly. 
An explicit computation shows 
 \be\label{deltaF}
   \delta_{\Lambda}F_{\mu\nu}{}^{M} \ = \ \widehat{\cal L}_{\Lambda}F_{\mu\nu}{}^{M}+\partial^{M}\big(\partial_{[\mu}\Lambda^N  
     A_{\nu]N}\big)\;.
 \ee
Thus, while $F_{\mu\nu}{}^{M}$ is not fully gauge covariant, 
by the section condition it is gauge invariant in terms with $F_{\mu\nu}{}^{M}\partial_{M}$. 
This will be sufficient for all its appearances in this paper. 
Similarly, one verifies that the general variation of the field strength $F_{\mu\nu}{}^M$
takes the form
 \be
  \delta F_{\mu\nu}{}^{M} \ = \ D_{\mu}(\delta A_{\nu}{}^{M})-D_{\nu}(\delta A_{\mu}{}^{M})
  +\partial^M(A_{[\mu}{}^{N}\delta A_{\nu]N})\;,
 \ee      
while its Bianchi identity is given by 
  \be\label{FBianchi}
  D_{[\mu}F_{\nu\rho]}{}^{M} \ = \ -\partial^M \Big(A_{[\mu}{}^{N}\partial_{\nu}A_{\rho]N}
  -\frac{1}{3}A_{[\mu N}\big[A_{\nu},A_{\rho]}\big]^N_{ C}\Big)\;.
 \ee
I.e.\ also all these relations are covariant up to terms that vanish under contraction 
with $\partial_{M}$ due to the section constraint. In the spirit of the tensor hierarchies
of gauged supergravity~\cite{deWit:2005hv,deWit:2008ta}, this suggests to introduce 
a 2-form potential $B_{\mu\nu}$ as
 \be\label{FHat}
  {\cal F}_{\mu\nu}{}^{M} \ \equiv \ F_{\mu\nu}{}^{M}-\partial^{M}B_{\mu\nu}\;,
 \ee
with proper transformation behavior, to compensate for the non-covariance, cf.~\cite{Hohm:2013new1}. For the actions discussed in this paper this extension will not be relevant, as the field strength always appears under contractions such that the non-covariant terms vanish.

\subsection{Gauge connection for $SL(2,\mathbb{R})$}
We now introduce the second gauge connection, $B_{\mu M}$, that
formally plays the role of an $SL(2,\mathbb{R})$ gauge field. 
As such, we will introduce covariant derivatives both with respect to
$A$ and $B$, which read on a general vector,  
  \be
   {\cal D}_{\mu}V_M \ = \ \partial_{\mu}V_M-A_{\mu}{}^{K}\partial_K V_{M}
    -\left(\partial_{M} A_{\mu}{}^{K}-\partial^KA_{\mu M}\right)V_{K}+B_{\mu}{}^{K}f_{KM}{}^{L}V_L\;.
    \label{cDV}
  \ee  
This is a  fully  
 covariant derivative, with respect to $\Lambda$ gauge transformations and local $SL(2,\mathbb{R})$ 
 transformations  
 with parameter $\Sigma_M$, provided $B$ transforms as 
  \be\label{shortdeltaB}
   \delta B_{\mu M} \ = \ {\cal D}_{\mu}\Sigma_M +\widehat{\cal L}_{\Lambda}B_{\mu M}\;,
  \ee
 where $\widehat{\cal L}_{\Lambda}$ acts on $B_{\mu M}$ as a vector, see (\ref{GennnLie}).  
Writing this out explicitly, we have  
  \be\label{BgaugeVar} 
 \begin{split}
  \delta B_{\mu M} \ &= \  \partial_{\mu}\Sigma_M-A_{\mu}{}^{K}\partial_K\Sigma_M
  -\big(\partial_M A_{\mu}{}^{K}-\partial^K A_{\mu M}\big)\Sigma_K 
  -\Sigma^K f_{KM}{}^{N} B_{\mu N}+ \widehat{\cal L}_{\Lambda}B_{\mu M}\;.
 \end{split}
 \ee
It is non-trivial that simultaneous $SL(2,\mathbb{R})$ and $\Lambda^M$ gauge transformations 
are consistent, in particular that they close. Closure can, however, be easily 
established using the result (\ref{epscov}) that the structure constants are $\Lambda$ gauge invariant:
 \be
 \begin{split}
  \big[\delta_{\Lambda},\delta_{\Sigma}\big]\,V^M \ &= \ 
  \delta_{\Lambda}\big(f^{MNK}\Sigma_{N}V_K\big)-\delta_{\Sigma}\big(\widehat{\cal L}_{\Lambda}V^M\big)\\
  \ &= \ f^{MNK}\Sigma_{N}\widehat{\cal L}_{\Lambda}V_K-\widehat{\cal L}_{\Lambda}\big(f^{MNK}\Sigma_{N}V_K\big) \\
  \ &= \ -f^{MNK}(\widehat{\cal L}_{\Lambda}\Sigma_{N})V_K \ \equiv \ \delta_{\Sigma'}V^M  \;, 
 \end{split} 
 \ee 
with the effective parameter $\Sigma_M'  =  -\widehat{\cal L}_{\Lambda}\Sigma_M$. 
Although we have closure,  
we will see that in the following there are not really two completely independent gauge symmetries 
with parameters $\Lambda_M$ and $\Sigma_M$. Rather, gauge invariance of the theory requires 
an extension of the section constraint (\ref{sectioncon}) involving field components 
of $A$ and $B$ (and correspondingly of their gauge parameters). 

In order to state these constraints 
it will be convenient to introduce the following combinations of $A$ and $B$
(and their parameters) 
\be
  \begin{split}
  \tilde{B}^M \ &= \ B^M-f^{MNK}\partial_N A_{K} \;, \\
  \tilde{\Sigma}^M \ &= \ \Sigma^M-f^{MNK}\partial_N \Lambda_{K}\;. 
  \end{split}
  \label{tilde}
 \ee
The reason is that in terms of these variables the complete version of the  section condition (\ref{sectioncon})
can be written most concisely (while the action and gauge transformations are more 
naturally written in terms of $B$). The full set of constraints for the following construction is
given by the requirement that 
 \be
  \mathbb{P}_{KL}{}^{MN} C_M\otimes C_M' \ = \ 0\;,  \qquad 
  \forall \; C, C' \ \in \ \{\partial , \tilde{B},\tilde{\Sigma}\}\;, 
  \label{fullconstraints}
 \ee 
with the projector from (\ref{projector}), and where $C$ and $C'$ denotes any elements of the list above. 
For instance, taking $C'_M=\tilde{B}_M$ and $C_M=\partial_M$, 
the constraint states that $\tilde{B}^M\partial_M=0$ in arbitrary combinations, 
in particular $\partial_M\tilde{B}^M=0$.
(Sometimes we leave out $\otimes$ when there is no possible confusion.) 
Another special case is 
\be\label{partialSigmaconstr}
 f^{MNK}\partial_N\otimes\tilde{\Sigma}_K \ = \ 0\;.
\ee
From this we 
can immediately derive some further constraints. Consider 
 \be
  0 \ = \  \tilde{B}^M \otimes \partial_M \ = \ (B^M-f^{MNK}\partial_NA_{K})\otimes \partial_M \ = \ B^M\otimes \partial_M \;,
 \ee
using in the last step $f^{MNK}\partial_N\otimes \partial_K=0$, which is implied by the 
constraint in (\ref{fullconstraints}).  The analogous conclusion follows for the gauge parameter $\Sigma_M$. 
Thus, in addition to $\tilde{B}^M\partial_M=\tilde{\Sigma}^M\partial_M=0$ the constraints also imply 
 \be\label{explBconstr}
  B^M \partial_M  \ = \ 0\;,\qquad \Sigma^M \partial_M  \ = \ 0\;. 
 \ee 
Another curious consequence follows by multiplying  $f^{MNK}\partial_N\otimes \partial_K=0$ 
with $f_{MPQ}$ and using standard identities for the structure constants (\ref{STRk}):
 \be\label{derconstr}
  \partial_P\otimes \partial_Q-\partial_Q\otimes \partial_P \ = \ 0\;.
 \ee
In other words, here the section constraints imply that the order of partial derivatives 
can be changed in arbitrary products.   Similarly, taking $C_M=\partial_M$ and $C'_M=\tilde{B}_M$
we obtain 
 \be\label{partialBant}
  \partial_M\otimes \tilde{B}_N - \partial_N\otimes \tilde{B}_M \ = \ 0\;.
 \ee
The analogous relation holds also for $\tilde{\Sigma}$.   
We stress that this relation does not hold for $B$.

Finally, we present an alternative form of the gauge transformations  of $B_{\mu M}$. 
The conventional form (\ref{shortdeltaB}) is fixed by the requirement that covariant 
derivatives transform  covariantly. In particular, $B_{\mu M}$ transforms as a 
vector under $\Lambda$ transformations. On the other hand, in the next section we will 
introduce a Chern-Simons action of the form $\int B_M\wedge F^M$, whose invariance 
requires $B$ to be a $\Lambda$ density of weight one rather than a vector.  Surprisingly, 
it turns out that as a consequence of the section constraints (\ref{fullconstraints}), the 
variation of $B$ can be rewritten so that a density term $\partial_N\Lambda^N$ 
appears. Specifically, we show that $\delta B$ can equivalently be written as 
\be\label{fullBgauge}
 \begin{split}
  \delta B_{\mu M} \ = \  &\;\partial_{\mu}\Sigma_M-A_{\mu}{}^{K}\partial_K\Sigma_M
  -\big(\partial_M A_{\mu}{}^{K}-\partial^K A_{\mu M}\big)\Sigma_K -\partial_KA_{\mu}{}^{K}\Sigma_M \\
  &+\widehat{\cal L}_{\Lambda}B_{\mu M}+\partial_N\Lambda^N B_{\mu M}\;. 
  \end{split}
 \ee 
This again takes the form of (\ref{shortdeltaB}), but now with $B$ and $\Sigma$ being 
$\Lambda$ densities (of weight one) not transforming under the local $SL(2,\mathbb{R})$ 
and with ${\cal D}_{\mu}$ and $\widehat{\cal L}_{\Lambda}$ acting accordingly.  
Therefore, in presence of a separate $SL(2,\mathbb{R})$ gauge symmetry, and with the 
section constraints (\ref{fullconstraints}), there is no 
invariant distinction between a $\Lambda$ vector and a vector-density,
which is crucial for the following construction.
For this to happen, it  is essential that we impose the section constraints (\ref{fullconstraints}) for the 
combination $\tilde{B}_{\mu\,M}$ from (\ref{tilde}), and not for the 
$SL(2,\mathbb{R})$ connection $B_{\mu\,M}$\,.

Let us now prove the equivalence of (\ref{BgaugeVar}) and (\ref{fullBgauge}), which 
requires  
 \be\label{bizarrelemma}
   -\Sigma^K f_{KM}{}^{N} B_{\mu N} \ = \ -\partial_KA_{\mu}{}^{K}\Sigma_M+\partial_N\Lambda^N B_{\mu M}\;. 
 \ee  
We start by computing for the left-hand side  
 \be
  \begin{split}
  ({\rm l.h.s.}) \ \equiv \  &-\Sigma^K f_{KM}{}^{N} B_{\mu N}  \ = \  -\left(\tilde{\Sigma}^K+f^{KPQ}\partial_P\Lambda_Q\right) f_{KM}{}^{N} 
    \left(\tilde{B}_{\mu N}+f_{NRS}\partial^R A_{\mu}{}^{S}\right) \\
    \ = \  &-f^{KPQ}f_{KM}{}^{N} \partial_P\Lambda_Q\tilde{B}_{\mu N} -\tilde{\Sigma}^K f_{KM}{}^{N}f_{NRS}\partial^R A_{\mu}{}^{S} \\
    &-f^{KPQ}f_{KM}{}^{N}f_{NRS}\partial^R A_{\mu}{}^{S}\partial_P\Lambda_Q\;.
  \end{split}
 \ee   
Here we set to zero the term of the form $f\tilde{\Sigma}\tilde{B}$, as it vanishes by the 
constraints (\ref{fullconstraints}). Next, we simplify the various contractions of structure constants, using the 
identity stated after (\ref{STRk}), 
 \be 
  \begin{split}
      ({\rm l.h.s.})  \ &= \ 
      \partial_M\Lambda^N \tilde{B}_{\mu N} -\tilde{\Sigma}^K\partial_M A_{\mu K}+f_{NRS}\partial^R A_{\mu}{}^{S}(\partial_M\Lambda^N-\partial^N\Lambda_M) \\
   \ &= \ \partial_N\Lambda^N \tilde{B}_{\mu M} -\tilde{\Sigma}_M\partial^K A_{\mu K}+f_{NRS}\partial^R A_{\mu}{}^{S}(\partial_M\Lambda^N-\partial^N\Lambda_M)\;. 
 \end{split}
 \ee
We omitted  terms with $\tilde{\Sigma}^K\partial_K$, etc., and we used (\ref{partialBant}), together with its analogue for 
$\tilde{\Sigma}$, in the second equation. Using (\ref{partialBant}) once more and translating everything back in $B,\Sigma$ basis 
we obtain 
 \be 
  \begin{split}
      ({\rm l.h.s.})    \ &= \  \partial_N\Lambda^N (B_{\mu M}-f_{MPQ}\partial^PA_{\mu}{}^{Q}) 
   -\partial_K A_{\mu}{}^{K}(\Sigma_M-f_{MPQ}\partial^P\Lambda^Q) \\
   &\quad \;+ f_{NRS}\partial^R A_{\mu}{}^{S}(\partial_M\Lambda^N-\partial^N\Lambda_M) \\
   \ &= \ \partial_N\Lambda^N B_{\mu M}-\partial_K A_{\mu}{}^{K}\Sigma_M\\
   &\quad \;   -f_{MPQ}\partial_N\Lambda^N\partial^PA_{\mu}{}^{Q}+f_{MPQ}\partial^P\Lambda^Q \partial_N A_{\mu}{}^{N} \\
   &\quad\; + f_{NPQ}\partial_M\Lambda^N\partial^P A_{\mu}{}^{Q}
   -f_{NPQ} \partial^N\Lambda_M\partial^P A_{\mu}{}^{Q}\;. 
  \end{split}
 \ee
The first line on the right-hand side of the final equality coincides with the required right-hand side of (\ref{bizarrelemma}).  
Thus, it remains to show that the last four terms are zero. Using the Schouten identity $0=f_{[MPQ}\,\partial_{N]}\Lambda^N$ 
and the section constraint $\partial_P\,\partial^P=0$ one can check that among these four terms the first and third combine into one, so that we obtain 
for them in total 
 \be
  \begin{split}
  &f_{MNP}\partial_Q\Lambda^N \partial^P A_{\mu}{}^{Q}+f_{MPQ}\partial^P\Lambda^Q \partial_N A_{\mu}{}^{N}
  -f_{NPQ} \partial^N\Lambda_M\partial^P A_{\mu}{}^{Q} \\
  & \ = \ f_{MNP}\partial^P\Lambda^N \partial_Q A_{\mu}{}^{Q}+f_{MPQ}\partial^P\Lambda^Q \partial_N A_{\mu}{}^{N}
  \ = \ 0\;, 
  \end{split}
 \ee     
where in the final step we used (\ref{derconstr}) in the first term and the section constraint in the last term. 
We therefore proved (\ref{bizarrelemma}) and thus the alternative form (\ref{fullBgauge}) of the gauge transformations. 
Let us note that along similar lines one may verify that the 
gauge variation (\ref{BgaugeVar}) is compatible with the constraints (\ref{fullconstraints}). 

Finally, we introduce the field strength associated to this gauge connection as
\bea
  G_{\mu\nu M} &\equiv&  D_{\mu}B_{\nu M}-D_{\nu}B_{\mu M}-f_{MNK} B_{\mu}{}^{N} B_{\nu}{}^{K}
\;,
\label{defG}
\eea
with $A_\mu$-covariantized derivatives from (\ref{covDerGen}),
such that
\be
  \big[ \cD_{\mu},\cD_{\nu}\big]V_M \ = \ -\widehat{\cal L}_{F_{\mu\nu}} V_M+G_{\mu\nu}{}^{K} f_{KM}{}^{L} V_L\;,
\ee
extending (\ref{commF}).
Upon using the Schouten identity and the constraints similar to the computation of (\ref{bizarrelemma}),
this field strength may be recast in the form
\bea
  G_{\mu\nu M} &=&
  D_{\mu}B_{\nu M}-D_{\nu}B_{\mu M}-2\left(\partial_N A_{[\mu}{}^N\right) B_{\nu]M}
  \;.
  \label{defG2}
\eea
Again, this shows that as a consequence of the particular form of the section constraints
(\ref{fullconstraints}), the field $B_\mu$ simultaneously plays the role of an $SL(2,\mathbb{R})$
connection and of an $SL(2,\mathbb{R})$ singlet with non-trivial $\Lambda$-weight.

\section{$(3+3)$-dimensional theory}
Using the techniques developed above, we introduce the $(3+3)$-dimensional formulation 
of $D=4$ Einstein gravity. The action consists of three main ingredients: a (generalized) Chern-Simons-matter 
Lagrangian, a covariantized Einstein-Hilbert term and a scalar potential. In the following three subsections 
we introduce these actions and prove their gauge invariance.

\subsection{Chern-Simons term and scalar kinetic term}
The Chern-Simons action is defined by   
 \be
  S_{\rm CS} \ = \ \int d^3x \, d^3Y \,\varepsilon^{\mu\nu\rho}B_{\mu M}F_{\nu\rho}{}^{M}\;,
 \ee
 up to a pre-factor that we shall neglect in this subsection.
We will now show that 
this action is invariant under local $\Lambda$ transformations in that the Lagrangian 
transforms into a total derivative. First note that the field strength transforms according to (\ref{deltaF}), 
which implies that upon contraction with $B_M$, as in the Chern-Simons term, it transforms covariantly 
thanks to the constraint (\ref{explBconstr}). Then the full $\Lambda$ invariance follows with the 
form of the gauge variation in (\ref{fullBgauge}) that treats $B$ as a $\Lambda$ density:
 \be
 \begin{split}
  \delta_{\Lambda}S_{\rm CS} 
  \ &= \ \int  d^3x \, d^3Y \,\varepsilon^{\mu\nu\rho}\left(\Lambda^N\partial_N(B_{\mu M}F_{\nu\rho}{}^{M})
  +\partial_N\Lambda^N B_{\mu M}F_{\nu\rho}{}^{M}\right) \\
  \ &= \ \int  d^3x \, d^3Y \,\varepsilon^{\mu\nu\rho}\,\partial_{N}\left(\Lambda^N B_{\mu M}F_{\nu\rho}{}^{M}\right) \ = \ 0\;, 
 \end{split}
 \ee 
where we used in the first line that the covariant terms in the variation of $B$ and $F$ combine into the 
Lie derivative of a scalar. 
 
Next, we turn to the invariance under local $SL(2,\mathbb{R})$ transformations parametrized by $\Sigma_M$. 
The gauge field $A$ and thus its field strength $F$ are inert under these transformations, while 
$\delta_{\Sigma} B_{\mu M}={\cal D}_{\mu}\Sigma_M$. Here we take again the form of the gauge variation 
in (\ref{fullBgauge}), so that the covariant derivative ${\cal D}_{\mu}$ acts on $\Sigma_M$ as a $\Lambda$ density. 
Consequently, we can integrate by parts with this covariant derivative and obtain for the gauge variation of the action 
 \be
  \delta_{\Sigma}S_{\rm CS} \ = \ \int d^3x \,d^3Y \,\varepsilon^{\mu\nu\rho}\,{\cal D}_{\mu}\Sigma_M F_{\nu\rho}{}^{M} 
  \ = \ -\int d^3x \,d^3Y \,\varepsilon^{\mu\nu\rho}\,\Sigma_M D_{\mu}F_{\nu\rho}{}^{M} \ = \ 0 \;, 
 \ee
using the Bianchi identity (\ref{FBianchi}) and the constraint (\ref{explBconstr}) in the last step. 
In total we have shown that the Chern-Simons term is invariant under all local symmetries 
except the $(2+1)$-dimensional diffeomorphisms parameterized by $\xi^{\mu}(x,Y)$, which will 
be discussed in the next section. 

Finally let us turn to the scalar kinetic term involving ${\cal M}_{MN}$, which 
transforms under the local symmetries as 
 \be
  \delta{\cal M}_{MN} \ = \ \widehat{\cal L}_{\Lambda}{\cal M}_{MN}-2\Sigma^P f_{P(M}{}^{Q} {\cal M}_{N)Q}\;.
 \ee  
Thus, the fully covariant derivative of ${\cal M}_{MN}$ reads 
 \be
  {\cal D}_{\mu}{\cal M}_{MN} \ = \ D_{\mu}{\cal M}_{MN}+2B_{\mu}{}^{P} f_{P(M}{}^{Q}{\cal M}_{N)Q}\;,
  \label{covDM}
 \ee
with the covariant derivative $D_{\mu}$ with respect to $A$  defined in (\ref{covM}). 
This derivative is manifestly covariant under local $\Lambda$ and $\Sigma$ transformations. 
For covariance under the latter symmetries we have to employ the original form (\ref{BgaugeVar}) of the 
gauge transformations that treats $B_{\mu M}$ as a conventional $SL(2,\mathbb{R})$ gauge field. 

Summarizing, we can define the total action consisting of scalar-kinetic term and Chern-Simons term, 
  \be\label{CSmatter}
  S_{\rm CS-matter} \ = \ \int d^3x \, d^3Y \Big(-\tfrac{1}{2\sqrt{2}}\varepsilon^{\mu\nu\rho}B_{\mu M}  F_{\nu\rho}{}^{M} 
  +\tfrac{1}{16}eg^{\mu\nu}
  {\cal D}_{\mu}{\cal M}^{MN}{\cal D}_{\nu}{\cal M}_{MN}\Big)\;, 
 \ee 
where we inserted the proper coefficient of the Chern-Simons term.  
This action is manifestly invariant under $\Lambda$ and $\Sigma$ gauge transformations.
Curiously, however, in order to make the $\Sigma$ invariance manifest we had to 
employ two different but equivalent forms of $\delta B$ for the scalar kinetic term 
and the Chern-Simons term.

\subsection{Covariantized Einstein-Hilbert term}
We next discuss the Einstein-Hilbert term in the `dreibein' formalism with $e_{\mu}{}^{a}$ 
and spin connection $\omega_{\mu}{}^{a}$, which we can treat as a Lorentz vector. 
Under local $\Lambda$ transformations they transform as  
\be\label{LambdaGravVar}
\begin{split}
 \delta_{\Lambda}e_{\mu}{}^{a} \ &= \ \Lambda^N\partial_N e_{\mu}{}^a+\partial_N\Lambda^N e_{\mu}{}^{a}\;, \\
 \delta_{\Lambda}\omega_{\mu}{}^{a} \ &= \ \Lambda^N\partial_N \omega_{\mu}{}^{a}\;,   
\end{split}
\ee
and so their covariant derivatives with respect to $A$ read 
 \be
 \begin{split}
  D_{\mu}e_{\nu}{}^{a} \ &= \ \partial_{\mu}e_{\nu}{}^{a}-A_{\mu}{}^{N}\partial_N e_{\nu}{}^{a}-\partial_N A_{\mu}{}^{N} e_{\nu}{}^{a}\;, \\
  D_{\mu}\omega_{\nu}{}^{a} \ &= \ \partial_{\mu}\omega_{\nu}{}^{a} -A_{\mu}{}^{N}\partial_N \omega_{\nu}{}^{a}\;. 
 \end{split}
 \label{covDe}
 \ee 
We can now write an `$A$ covariantization' of the $D=3$ Einstein-Hilbert term, 
\be\label{firstEinsteinH}
\begin{split}
 S_{\rm EH}  \ &= \ \int d^3x \, d^3Y\,  e\,R  \ = \ -\int d^3x \, d^3Y\, \varepsilon^{\mu\nu\rho}\,
  e_{\mu}{}^a R_{\nu\rho a} \\
  \ &\equiv \ - \int d^3x \, d^3Y\, \varepsilon^{\mu\nu\rho}\,
  e_{\mu}{}^a
   \big(D_{\nu}\omega_{\rho a}-D_{\rho}\omega_{\nu a}+\varepsilon_{abc}
   \omega_{\nu}{}^b\omega_{\rho}{}^c\big)\;, 
\end{split}
\ee
where we defined the covariantized $D=3$ Riemann tensor and 
the corresponding Ricci scalar. This action 
is manifestly invariant under local $\Lambda$ transformations, because $e_{\mu}{}^{a}$
transforms as a $\Lambda$ density so that  the full Lagrangian transforms into a 
total derivative. 

In the action (\ref{firstEinsteinH}) we may treat the spin connection $\omega_{\mu}{}^{a}$
as an independent field or as determined by means of its field equations in terms of 
(derivatives of) the dreibein $e_{\mu}{}^{a}$. More precisely,  as in standard gravity the 
field equation for $\omega_{\mu}{}^{a}$ implies vanishing torsion,  
 \be
  T_{\mu\nu}{}^{a} \ = \ D_{\mu}e_{\nu}{}^{a}-D_{\nu}e_{\mu}{}^{a} +\varepsilon^{abc}e_{\mu b}\,\omega_{\nu c}
  -\varepsilon^{abc}e_{\nu b}\, \omega_{\mu c} \ = \ 0\;.
 \ee 
This can be solved in the standard fashion, giving $\omega=\omega(e,A)$, the only 
difference being that all occuring derivatives are covariant with respect to $A$. 
Specifically, the Lorentz vector spin connection is related to the usual one via 
$\omega_{\mu}{}^{ab} = -\epsilon^{abc} \omega_{\mu c}$, which in turn is given by 
 \be\label{explspinconn}
  \omega_{abc} \ = \ \frac{1}{2}\left(\Omega_{abc}  -\Omega_{bca}  +\Omega_{cab}  \right)\;, \qquad 
  \Omega_{abc} \ = \ -\Omega_{bac} \ = \ (e_{a}{}^{\mu} e_{b}{}^{\nu}-e_{b}{}^{\mu} e_{a}{}^{\nu}) D_{\mu}e_{\nu c}\;, 
 \ee
where all indices haven been flattened.   
For definiteness we view $\omega$ as determined in this way. 

We now turn to the local Lorentz transformations with parameter $\lambda_a$, 
 \be\label{localLorentz}
   \delta_{\lambda}e_{\mu}{}^{a} \ = \ \varepsilon^{abc}e_{\mu b}\lambda_c\;, \qquad
   \delta_{\lambda}\omega_{\mu}{}^{a}\ = \ D_{\mu}(\omega)\lambda^a 
   \ \equiv  \ D_{\mu}\lambda^a+\varepsilon^{abc}\, \omega_{\mu b}\,\lambda_c\;,
 \ee  
where we indicated by $D_{\mu}(\omega)$ the covariant derivative with respect to 
both $\omega$ and $A$. It turns out that due to the $A$ covariantization of 
the Riemann tensor it no longer transforms fully covariantly under 
local Lorentz transformations.  In order to see this we compute 
  \be
   \delta_{\lambda}R_{\nu\rho\, a} \ = \ 2D_{[\nu}\,\delta \omega_{\rho ]a} \ = \ \big[D_{\nu},D_{\rho}\big]\lambda_a\;. 
  \ee 
Since the covariant derivative denotes the full covariant derivative with respect to both the spin connection and 
with respect to $A_{\mu}$, the commutator does not only give the Riemann tensor, which 
represents the covariant term, but also the curvature $F$ of $A$. Therefore, 
denoting the non-covariant variation by $\Delta^{\rm nc}$ we find 
 \be
  \Delta^{\rm nc}_{\lambda} R_{\nu\rho\, a} \ = \ -F_{\nu\rho}{}^{M}\partial_M\lambda_a\;.
 \ee 
The Einstein-Hilbert term then transforms as 
 \be\label{nonivEinstein}
  \delta_{\lambda} \big(-\varepsilon^{\mu\nu\rho}e_{\mu}{}^{a}R_{\nu\rho\,a}\big) \ = \ \varepsilon^{\mu\nu\rho} e_{\mu}{}^{a} F_{\nu\rho}{}^{M}\partial_M\lambda_a\;.
 \ee 
This non-invariance can be cured by introducing an improved Riemann tensor  
  \be
  \widehat{R}_{\mu\nu\,a} \ = \  R_{\mu\nu\,a} +\frac{1}{2}e\varepsilon_{\rho\sigma[\mu} F^{\rho\sigma M}\partial_M e_{\nu] a}\;, 
 \ee 
which leads to the following modification of the Einstein-Hilbert term,
 \be\label{improvedEH}
  e\widehat{R} \ = \ -\varepsilon^{\mu\nu\rho}e_{\mu}{}^{a}\widehat{R}_{\nu\rho\,a} \ = \ 
  eR-ee^{a\mu}e^{b\nu} F_{\mu\nu}{}^{M} e_{b}{}^{\rho}\partial_M e_{\rho a}\;.
 \ee
The new term induces a non-covariant variation under the local Lorentz transformations (\ref{localLorentz})
due to the $\partial_M$ derivative: 
 \be
 \begin{split}
  \delta_{\lambda}\big(-ee^{a\mu}e^{b\nu} F_{\mu\nu}{}^{M} e_{b}{}^{\rho}\partial_M e_{\rho a}\big)
  \ &= \ -e e^{a\mu} e^{b\nu} F_{\mu\nu}{}^{M} e_{b}{}^{\rho} \varepsilon_{a}{}^{cd} e_{\rho c} \partial_M\lambda_d  \\
  \ &= \ -e e^{a\mu} e^{b\nu} \varepsilon_{ab}{}^{d}F_{\mu\nu}{}^{M}  \partial_M\lambda_d \\
  \ &= \  -\varepsilon^{\mu\nu\rho}e_{\mu}{}^{a}F_{\nu\rho}{}^{M}  \partial_M\lambda_a\;.  
 \end{split}
 \ee
This cancels exactly  (\ref{nonivEinstein}) and so the improved Einstein-Hilbert term is 
invariant under local Lorentz transformations. Moreover, 
it is still invariant under $\Lambda$ transformations, although this is not totally 
trivial due  the $\partial_M e$ term.  From (\ref{LambdaGravVar}) 
we find, however, 
 \be
  \delta_{\Lambda}\big(e_{b}{}^{\rho}\partial_M e_{\rho a}\big) \ = \ \widehat{\cal L}_{\Lambda}\big(e_{b}{}^{\rho}\partial_M e_{\rho a}\big)
  +e_{b}{}^{\rho}e_{\rho a} \partial_M\partial_N\Lambda^N \;, 
 \ee
so that the second, non-covariant term is symmetric in $a,b$ and hence drops out   
from (\ref{improvedEH}), where this is contracted with the antisymmetric $F^{ab}$. 
Summarizing, the improved Einstein-Hilbert term (\ref{improvedEH}) is invariant under local 
Lorentz and $\Lambda$ transformations.

\subsection{Scalar potential}
In this subsection we prove that the potential term (\ref{scalarIntro}), 
 \be\label{scalar}
 \begin{split}
  eV({\cal M},g)& \ = \ -\frac{3}{16}e\Big( {\cal M}^{KL}\partial_K {\cal M}^{MN}\partial_L {\cal M}_{MN}
  -4{\cal M}^{MN}\partial_M {\cal M}^{PQ}\partial_Q {\cal M}_{PN}\Big) \\
  &-\frac{1}{2}e g^{-1}\partial_M g\, \partial_N{\cal M}^{MN}-\frac{1}{4}e {\cal M}^{MN}g^{-1}\partial_Mg\,g^{-1}\partial_Ng
  -\frac{1}{4}e  {\cal M}^{MN}\partial_M g^{\mu\nu}\,\partial_N g_{\mu\nu}\;, 
 \end{split}
 \ee 
is gauge invariant under $\Lambda$ and $\Sigma$ transformations. 
At first sight one would suspect that the proof of $\Lambda$ invariance proceeds 
more or less precisely as in DFT, with the dreibein $e_{\mu}{}^{a}$ and its determinant 
$e\equiv \det(e_{\mu}{}^{a})\equiv \sqrt{|\det g_{\mu\nu}|}\equiv \sqrt{g}$ playing the role of the dilaton density in string theory. From 
(\ref{LambdaGravVar}) we infer that $e_{\mu}{}^{a}$ indeed  transforms as a 
$\Lambda$ density of weight one. However, this implies that $e$ transforms 
as a density of weight $3$, which is puzzling because with ${\cal M}_{MN}$ being a tensor 
and not a tensor density, invariance of (\ref{scalar}) seems to require $e$ to have weight one. 
A related puzzle is that we require the separate local $SL(2,\mathbb{R})$ symmetry, and due to 
the partial derivatives in (\ref{scalar}) it appears challenging to make the action invariant. 
The resolution of both obstacles is related and again hinges on the particular form of the constraints 
(\ref{fullconstraints}), which imply a relation between the $\Lambda$ and $\Sigma$
parameters. These will lead to additional density-type terms $\partial_N\Lambda^N$ in the variation, 
which in turn complete the weight of the Lagrangian to the `correct' one. 
We then establish full invariance of the potential term.  

We now turn to a detailed computation of the gauge variation of (\ref{scalar}), 
starting with the local $SL(2,\mathbb{R})$ symmetry. 
We first recall that the scalars transform under $\Sigma$ as 
 \be
  \delta_{\Sigma}{\cal M}_{MN} \ = \ -\Sigma^P f_{PM}{}^{Q} {\cal M}_{QN}-\Sigma^P f_{PN}{}^{Q} {\cal M}_{MQ}\;.
 \ee 
Let us first compute the gauge variation of the partial derivative $\partial_K {\cal M}_{MN}$, which 
contains covariant and non-covariant terms. The covariant terms automatically cancel out 
in the variation of the potential, the latter being an $SL(2,\mathbb{R})$ singlet.   
Thus, we collect only the non-covariant terms, denoting the corresponding 
variation by $\Delta^{\rm nc}$, 
 \be\label{delpartM}
 \begin{split}
  \Delta^{{\rm nc}}_{\Sigma}(\partial_K {\cal M}_{MN}) \ &= \ \Sigma^{P} f_{PK}{}^{Q}\partial_{Q}{\cal M}_{MN}
  -2\partial_K\Sigma^P\, f_{P(M}{}^{Q} {\cal M}_{N)Q} \\
  \ &= \ f^{PRS}\partial_R\Lambda_S  f_{PK}{}^{Q}\partial_{Q}{\cal M}_{MN}
  -2\partial_K\Sigma^P\, f_{P(M}{}^{Q} {\cal M}_{N)Q} \\
  \ &= \ -\partial_P\Lambda^P\,\partial_K {\cal M}_{MN}-2\partial_K\Sigma^P\, f_{P(M}{}^{Q} {\cal M}_{N)Q} \;. 
 \end{split}
 \ee
In the first line we used that the non-covariant terms are those where $\partial_K$
acts on the gauge parameter, while the first term compensates for $\partial_Q$
being inert under $SL(2,\mathbb{R})$.  In the first term of the second line we expressed $\Sigma$ in 
terms of $\tilde{\Sigma}$ and used the constraint (\ref{partialSigmaconstr}). 
The first term in the last line then shows that $\partial{\cal M}$ receives
 a weight $-1$. This is precisely the weight needed for invariance: since 
 the terms in the first line of (\ref{scalar}) have two $\partial{\cal M}$, each of weight $-1$, 
they combine with the $e$ of weight $3$ to a total weight of $1$. 
Rewriting the second term in the last line of (\ref{delpartM}) in terms of $\tilde{\Sigma}$ we get 
 \be
  \begin{split}
  -2\partial_K\Sigma^P\, f_{P(M}{}^{Q} {\cal M}_{N)Q}  \ &= \ 
  -2\partial_K(\tilde{\Sigma}^P+f^{PRS}\partial_R\Lambda_S)f_{P(M}{}^{Q} {\cal M}_{N)Q} \\
  \ &= \ -2\partial_K\tilde{\Sigma}^Pf_{P(M}{}^{Q} {\cal M}_{N)Q}+2\partial_K(\partial_{(M}\Lambda^Q-\partial^Q\Lambda_{(M}){\cal M}_{N)Q}\;. 
  \end{split}
 \ee
Interestingly, the second term coincides with the non-covariant variation of $\partial_K{\cal M}_{MN}$ under 
$\Lambda$ transformations. More precisely, defining the non-covariant variation $\Delta^{\rm nc}_{\Lambda}=\delta_{\Lambda}-\widehat{\cal L}_{\Lambda}$
one finds  
  \be
  \Delta^{\rm nc}_{\Lambda}(\partial_K {\cal M}_{MN}) \ = \ 2\partial_K(\partial_{(M}\Lambda^Q-\partial^Q\Lambda_{(M}){\cal M}_{N)Q}\;. 
 \ee
We have therefore shown 
 \be\label{DelSigmaM}
   \Delta^{{\rm nc}}_{\Sigma}(\partial_K {\cal M}_{MN}) \ = \  -2\partial_K\tilde{\Sigma}^Pf_{P(M}{}^{Q} {\cal M}_{N)Q}
   -\partial_P\Lambda^P\,\partial_K {\cal M}_{MN}+ \Delta^{\rm nc}_{\Lambda}(\partial_K {\cal M}_{MN})\;. 
 \ee
It is this form that we will use below to verify invariance of the full potential. 

In order to compute the full variation of the potential (\ref{scalar}) we need the variation 
of $g_{\mu\nu}$ and $g=|\det g_{\mu\nu}|$. 
Since $g_{\mu\nu}$ is inert under local $\Sigma$ transformations, the only non-covariant variation of 
$\partial_M g_{\mu\nu}$ originates by the partial derivative not rotating under $SL(2,\mathbb{R})$. Thus,  
 \be\label{DelTASIGMAg}
  \Delta_{\Sigma}^{\rm nc}(\partial_M g_{\mu\nu}) \ = \ \Sigma^P f_{PM}{}^{Q} \partial_Q g_{\mu\nu}
  \ = \ -\partial_P\Lambda^P \partial_M g_{\mu\nu}\;, 
 \ee
where the last step follows by precisely the same argument as in (\ref{delpartM}).   
Thus, as for $\partial{\cal M}$, this gives a weight $-1$ to $\partial_M g$, so that 
with the determinant $e$ having weight $+3$ 
this completes the weight of the terms in the second and third line of (\ref{scalar}) to the desired $+1$. 
Note that there is no $\tilde{\Sigma}$ term left in (\ref{DelTASIGMAg}). 

We have written the variations of the various terms with $\Lambda$ and $\tilde{\Sigma}$. 
Our strategy is now to check cancellation of terms with $\Lambda$ and $\tilde{\Sigma}$ separately, 
starting with the $\tilde{\Sigma}$ invariance. We first note from (\ref{DelSigmaM}) 
 \be
  \Delta_{\tilde{\Sigma}}^{\rm nc}(\partial_N{\cal M}^{MN}) \ = \ -\partial_N \tilde{\Sigma}^Q f_{Q}{}^{MP}{\cal M}_{P}{}^{N}\;, 
 \ee
where we used that by the section constraint (\ref{partialSigmaconstr}) one term is zero. 
Again by the section constraint this  
vanishes when contracted with $\partial_M g$.  Thus, for the non-covariant $\tilde{\Sigma}$ terms  
it remains to verify cancellation in 
the first line of (\ref{scalar}). Denoting these two terms in the potential as 
$-\frac{3}{16}(V^{(1)}+V^{(2)})+\cdots$ we compute with (\ref{DelSigmaM}) 
 \be\label{FIRStterm}
  \delta_{\tilde{\Sigma}} V^{(1)} \ = \ -4 {\cal M}^{KL} \partial_K \tilde{\Sigma}^Q f_{Q}{}^{MP} {\cal M}_{P}{}^{N} \partial_L {\cal M}_{MN}\;, 
 \ee 
and for the second term 
 \be 
 \begin{split}
  \delta_{\tilde{\Sigma}} V^{(2)} \ = \ &4 {\cal M}^{MN}\big(\partial_M\tilde{\Sigma}^R f_{R}{}^{PS} {\cal M}_{S}{}^{Q}+\partial_M\tilde{\Sigma}^R f_{R}{}^{QS} {\cal M}_{S}{}^{P}\big)
  \partial_Q {\cal M}_{PN}\\
  &+4{\cal M}^{MN} \partial_{M}{\cal M}^{PQ}\big(\partial_Q \tilde{\Sigma}^R f_{RP}{}^{S} {\cal M}_{SN}+
  \partial_Q\tilde{\Sigma}^R f_{RN}{}^{S} {\cal M}_{SP}\big)\;. 
 \end{split}
 \ee 
Distributing the terms this reads  
 \be 
 \begin{split}
   \delta_{\tilde{\Sigma}} V^{(2)} \ = \  \;&4 {\cal M}^{MN}\partial_M\tilde{\Sigma}^R f_{R}{}^{PS} {\cal M}_{S}{}^{Q}\partial_Q {\cal M}_{PN}
  +4 {\cal M}^{MN}\partial_M\tilde{\Sigma}^R f_{R}{}^{QS} {\cal M}_{S}{}^{P}
  \partial_Q {\cal M}_{PN}\\
  &+4\partial_{M}{\cal M}^{PQ} \partial_Q \tilde{\Sigma}^R f_{RP}{}^{M} +
  4{\cal M}^{MN} \partial_{M}{\cal M}^{PQ}\partial_Q\tilde{\Sigma}^R f_{RN}{}^{S} {\cal M}_{SP}\\
   \ = \  \;& 4 {\cal M}^{MN}\partial_M\tilde{\Sigma}^R f_{R}{}^{PS} {\cal M}_{S}{}^{Q}\partial_Q {\cal M}_{PN}+
  4{\cal M}^{MN} \partial_{M}{\cal M}^{PQ}\partial_Q\tilde{\Sigma}^R f_{RN}{}^{S} {\cal M}_{SP}\;. 
 \end{split}
 \ee  
Here we used that the second and third term in the first equation are zero by the constraint.
Next we use a Schouten identity in the second term of the last line, with a total antisymmetrization 
in $Q,R,N,S$ in the $\partial\tilde{\Sigma}f$ term. One term vanishes by the section constraint
and we obtain  
 \be
  \begin{split}
   \delta_{\tilde{\Sigma}} V^{(2)} &\ = \  4 {\cal M}^{MN}\partial_M\tilde{\Sigma}^R f_{R}{}^{PS} {\cal M}_{S}{}^{Q}\partial_Q {\cal M}_{PN}
  +4{\cal M}^{MN} \partial_{M}{\cal M}^{PQ}\partial_N\tilde{\Sigma}^R f_{RQ}{}^{S} {\cal M}_{SP} \\
  &\qquad +4{\cal M}^{MN} \partial_{M}{\cal M}^{PQ}\partial^S\tilde{\Sigma}^R f_{RNQ} {\cal M}_{SP} \\
  & \ = \ 4 {\cal M}^{MN}\partial_M\tilde{\Sigma}^R f_{R}{}^{PS} {\cal M}_{S}{}^{Q}\partial_Q {\cal M}_{PN}
  +4{\cal M}^{MN} \partial_{M}{\cal M}^{PQ}\partial_N\tilde{\Sigma}^R f_{RQ}{}^{S} {\cal M}_{SP} \\
  &\qquad +4{\cal M}^{QS} \partial_{Q}{\cal M}^{NP}\partial^M\tilde{\Sigma}^R f_{RSP} {\cal M}_{MN}\\
  & \ = \ 4{\cal M}^{MN} \partial_{M}{\cal M}^{PQ}\partial_N\tilde{\Sigma}^R f_{RQ}{}^{S} {\cal M}_{SP}\;. 
  \end{split}
 \ee 
Here we relabeled indices in the second equation in order to make it manifest that the first and third term cancel. 
The remaining term cancels against (\ref{FIRStterm}), completing the 
proof that the potential is $\tilde{\Sigma}$ invariant. 

Let us now turn to the $\Lambda$ invariance. Recall from (\ref{LambdaGravVar}) that 
 \be
  \begin{split}
    \delta_{\Lambda}g_{\mu\nu} \ &= \ \Lambda^N\partial_N g_{\mu\nu}+2\partial_N\Lambda^N g_{\mu\nu}\;, 
  \end{split}
 \ee   
which yields for the non-covariant $\Lambda$ variations\footnote{We note that the non-covariant variation of 
the last two terms in (\ref{scalar}) are equal. Therefore, $\Lambda$ gauge invariance does not 
determine their relative coefficients.} 
\be\label{DeltaPartg}
 \begin{split}
  g^{-1} \Delta^{\rm nc}_{\Lambda}(\partial_M g) \ &= \ 6\partial_M\partial_N\Lambda^N \;,  \\
  \Delta^{\rm nc}_{\Lambda}(\partial_M g_{\mu\nu}) \ &= \ 2\partial_M\partial_N\Lambda^N g_{\mu\nu}\;. 
 \end{split}
 \ee 
We can now use a result from DFT since the first line in the potential (\ref{scalar}) precisely agrees, 
up to the overall factor, with the corresponding terms in the DFT scalar curvature. We have to remember, however, 
not only to collect the non-covariant $\Lambda$ variations of $\partial{\cal M}$ terms, but also 
the same terms that originated from the $\Sigma$ variation above, see (\ref{DelSigmaM}).
In other words, each of the $\Delta^{\rm nc}_{\Lambda}$ terms gets doubled. 
Taking this factor of 2 into account we can read off the variation of the first line 
from eq.~(4.47) in \cite{Hohm:2010pp}   
  \be
  \delta_{\Lambda}\big(e{\cal M}^{KL}\partial_K {\cal M}^{MN}\partial_L {\cal M}_{MN}
  -4e{\cal M}^{MN}\partial_M {\cal M}^{PQ}\partial_Q {\cal M}_{PN}\big)  
  \ = \ -16e \partial_M\partial_N\Lambda^P\partial_P{\cal M}^{MN}\;. 
 \ee 
From (\ref{DeltaPartg}) we find for the variation of the second line 
 \be
  \begin{split}
  &\delta_{\Lambda}\Big(-\frac{1}{2}e g^{-1}\partial_M g\, \partial_N{\cal M}^{MN}-\frac{1}{4} e{\cal M}^{MN}g^{-1}\partial_Mg\,g^{-1}\partial_Ng
  -\frac{1}{4} e {\cal M}^{MN}\partial_M g^{\mu\nu}\,\partial_N g_{\mu\nu}\Big) \\
  & \ = \  -3 e \partial_M\partial_P \Lambda^P\, \partial_N{\cal M}^{NM} + e g^{-1}\partial_M g\,\partial_P\partial_Q\Lambda^M
  {\cal M}^{PQ} -e{\cal M}^{MN}g^{-1}\partial_Mg\,\partial_N\partial_P\Lambda^P\;. 
  \end{split}
 \ee  
The total variation of the potential is then given by 
 \be
 \begin{split}
  \delta_{\Lambda}(eV) \ = \ &\;3e\partial_M\partial_N\Lambda^P\partial_P{\cal M}^{MN}-3e\partial_M\partial_P\Lambda^P\partial_N{\cal M}^{MN} \\
  &+2\partial_Me\,\partial_P\partial_Q\Lambda^M{\cal M}^{PQ}-2{\cal M}^{MN}\partial_Me\,\partial_N\partial_P\Lambda^P\;.
 \end{split}
 \ee 
Next, we integrate by parts in the second line in order to remove $\partial_Me$ terms. 
The generated $\partial^3\Lambda$ terms cancel each other, while the remaining terms combine 
with those in the first line, so that 
 \be
   \delta_{\Lambda}(eV) \ = \ e\partial_M\partial_N\Lambda^P\partial_P{\cal M}^{MN}-e\partial_M\partial_P\Lambda^P\partial_N{\cal M}^{MN}\;.
 \ee  
Finally, using the section constraint in the form (\ref{derconstr}) to exchange $\partial_N$ and $\partial_P$ 
we see that the remaining two terms cancel. We have thus proved the complete gauge invariance of the 
potential.

\section{$(2+1)$-dimensional diffeomorphisms}

The full $(3+3)$-dimensional action that we have been putting together in the last section
takes the form
 \be\label{visionaryAction}
  S \ = \ \int d^3x \, d^3Y \Big(e\,\widehat{R}(e,\omega,A)-\tfrac1{2\sqrt{2}}\,\varepsilon^{\mu\nu\rho}B_{\mu M}  F_{\nu\rho}{}^{M} 
  +\tfrac{1}{16}eg^{\mu\nu}
  {\cal D}_{\mu}{\cal M}^{MN}{\cal D}_{\nu}{\cal M}_{MN} -e\,V({\cal M},g) \Big)\,.
 \ee 
 The first term is the covariantized Einstein-Hilbert term from (\ref{improvedEH}), the last term is the potential (\ref{scalar}), and the kinetic term carries the full covariant derivatives ${\cal D}_{\mu}$ from (\ref{covDM}). In the last section, we have shown that separately all terms are invariant under $\Lambda$ and $\Sigma$ gauge transformations.
 
Invariance of (\ref{visionaryAction}) 
under standard $x$-dependent $(2+1)$-dimensional diffeomorphisms is manifest.
In this section, we will discuss invariance of the action under those $(2+1)$-dimensional 
diffeomorphisms whose parameter $\xi^\mu$ also depends on the extra coordinates $Y$,
which turns out to be much more involved. This requires the interplay and various conspiracies among the variations of the four terms in (\ref{visionaryAction}), none of which is separately invariant. In particular, this 
generalised diffeomorphism invariance uniquely fixes all the relative coefficients in the action above.
For transparency of the presentation, we shall in the following discussion of invariance 
suppress a class of terms that cancel independently. These are of the
form ${\cal M}^{MN}\partial_M \otimes \partial_N$ with no other scalar field dependence than the single matrix ${\cal M}^{MN}$.
These terms cancel separately among themselves, and with 
the explicit parametrization~(\ref{VV}) adopted in the next section, it is straightforward to verify
that their cancellation is completely parallel to the calculation that ensures
standard diffeomorphism invariance in four-dimensional Einstein gravity. 
In particular, these terms do not interfere with the non-trivial checks of 
generalized diffeomorphism invariance that we present
in the following, giving rise to the cancellations of all the remaining structures.
Similarly, in the following we will also neglect all terms in the variation that carry explicit 
gauge fields. Such terms e.g.\ arise from the connection part upon partial integration 
from the fact that the integrand is not of $\Lambda$-weight one. 
Their vanishing can be shown by a separate calculation 
similar to establishing the $\Lambda$-invariance of the action in the last section.

The action of the gauge covariant diffeomorphisms on the scalars and the vielbein is expected to take the standard form
 \be\label{xidiff}
  \delta_{\xi}{\cal M}_{MN} \ = \ \xi^{\mu} {\cal D}_{\mu}{\cal M}_{MN}\;, \qquad
  \delta_{\xi}e_{\mu}{}^{a} \ = \ \xi^{\rho}D_{\rho} e_{\mu}{}^{a}+D_{\mu}\xi^{\rho} e_{\rho}{}^{a}\;,
 \ee
of a combined diffeomorphism and $(\Lambda$, $\Sigma)$ gauge transformation.
Accordingly, the covariant derivatives carry the connections from (\ref{covDM}), (\ref{covDe}).
In contrast, their action on the vector fields $A_{\mu}{}^{M}$, $B_{\mu}{}_{M}$
turns out to carry explicitly non-covariant terms. For their transformation laws, we start from the following ansatz
\bea
 \delta^{(0)}_{\xi} A_{\mu}{}^{M} &=& 
 \xi^{\nu}F_{\nu\mu}{}^{M}+{\cal M}^{MN}g_{\mu\nu}\partial_{N}\xi^{\nu}\;,
 \label{deltaxiA}\\[1.5ex]
\delta^{(0)}_{\xi} B_{\mu\,M} &=& \xi^{\nu}G_{\nu\mu\,M}+
f_{MN}{}^K\,F_{\mu\nu}{}^N\partial_K\xi^\nu
+f_{MNK}\partial^N
({\cal M}^{KL}g_{\mu\nu}\partial_{L}\xi^{\nu})
\label{varB}\\
&&{}
+\frac12\,
({\cal M}_{QN}\partial_M{\cal M}^{PN}) f^{QK}{}_P\partial_K\xi^\lambda g_{\lambda\mu}
-\sqrt{2}\, g_{\mu\lambda} \, e^{-1}\varepsilon^{\lambda\nu\rho}
\,{\cal D}_\nu\left(\partial_M \xi^\sigma g_{\rho\sigma} \right)
\;,\nonumber
\eea
which combines the covariant part of the transformation expressed in terms of the
field strengths from (\ref{fieldstr}), (\ref{defG}) with explicitly non-covariant terms
that are required for invariance of the action.
Their presence is already observed in the corresponding $(3+1)$-dimensional 
reformulation of four-dimensional Einstein gravity, that we review in the next section, 
cf.~(\ref{varA4}), (\ref{varB4}) below. 
As we will witness in the course of the calculation, both transformation laws will acquire yet further
(on-shell vanishing) contributions. We note, that the transformation law (\ref{varB}) is compatible with the
constraints (\ref{fullconstraints}) imposed on the vector field $B_{\mu\,M}$, as can be verified by a quick explicit computation.

Since the variation of the vector fields (\ref{deltaxiA}), (\ref{varB}) plays the crucial role in showing invariance 
of the action under generalized diffeomorphisms, let us first spell out the general variation of the Lagrangian
under variation of the vector fields, which up to total derivatives takes the form
\bea
\delta{\cal L} &=& 
 -\frac1{2\sqrt{2}} \,
\varepsilon^{\mu\nu\rho}\left( {\cal E}^{(A)}_{\nu\rho\vphantom{M}}{}^M\,\delta B_{\mu\,M}
+{\cal E}^{(B)}_{\nu\rho\,M}\,\delta A_{\mu}{}^{M} \right)
\;,
\label{varLAB}
\eea
with the combinations
\bea
{\cal E}^{(A)}_{\mu\nu\vphantom{M}}{}^M \hspace{-0.3em}&\equiv& \hspace{-0.3em}
F_{\mu\nu}{}^{M}
-\frac1{2\sqrt{2}} \,e\varepsilon_{\mu\nu\rho} \, f^{MK}{}_{L}
{\cal D}^\rho {\cal M}^{LN} {\cal M}_{NK} 
\;,\nonumber\\[1ex]
{\cal E}^{(B)}_{\mu\nu\,M} \hspace{-0.3em}&\equiv& \hspace{-0.3em}
G_{\mu\nu\,M}+\frac1{\sqrt{2}}\,\varepsilon_{\mu\nu\rho}\,\Big(
\partial_K\left(e{\cal M}_{ML}\,{\cal D}^\rho {\cal M}^{LK}\right)
-\frac1{4}\,e
\partial_M {\cal M}_{KL}\,{\cal D}^\rho {\cal M}^{KL}\Big)
+ \Omega_{\mu\nu\,M}
\,,
\label{duality}
\eea
exhibiting the duality equations relating vector and scalar fields,
typical in three dimensions.\footnote{
Strictly speaking, not all components of ${\cal E}^{(A)}_{\mu\nu\vphantom{M}}{}^M$
are independent equations of motion, since the vector field $B_{\mu\,M}$ is subject
to the constraints~(\ref{fullconstraints}), but this does not affect the proof of gauge invariance here.}
The term $\Omega_{\mu\nu\,M}$ comprises all contributions that descend from variation of the improved Einstein-Hilbert term,
whose explicit form will not be needed in the following. Note though that all these terms carry an explicit $\partial_M$
and thus vanish when contracted with another $\partial^M$.

Let us now study the variation of the action (\ref{visionaryAction}) under the generalized diffeomorphisms (\ref{xidiff})--(\ref{varB}).
First, we consider the covariantized 
Einstein-Hilbert term. In addition to the above listed fields, this term depends on the spin connection 
that transforms exactly like $e_{\mu}{}^{a}$ together with non-covariant terms descending from (\ref{deltaxiA})
\bea
\delta_\xi^{(0)}\omega_{\mu}{}^{ab} &=& {\cal L}_{\xi}(\omega_{\mu}{}^{ab})
-3 g_{\mu\nu}\,{\cal M}^{MN}\,e^{a\rho} \partial_M e_{\rho}{}^b\,\partial_N \xi^\nu
\;, 
\eea
which may be verified with (\ref{explspinconn}). 
However, the non-covariant term in this variation
is of the type ${\cal M}^{MN}\partial_M \otimes \partial_N$
that cancel separately. We note that the antisymmetric 
$\varepsilon^{\mu\nu\rho}$ satisfies
 \be
  \delta_{\xi}\varepsilon^{\mu\nu\rho} \ = \ 0 \ = \ {\cal L}_{\xi}\varepsilon^{\mu\nu\rho}+D_{\lambda}\xi^{\lambda}\,\varepsilon^{\mu\nu\rho}\,
\;,
 \ee
where now
 \be
  D_{\mu}\xi^{\nu} \ = \ \partial_{\mu}\xi^{\nu}-A_{\mu}{}^{N}\partial_{N}\xi^{\nu}
  \;.
 \ee   
Thus, as for standard diffeomorphisms, this object
is a density and the Einstein-Hilbert term is invariant except for terms that originate from 
the non-commutativity of $D_{\mu}$ and the non-covariant contributions from (\ref{deltaxiA}). 
Projected with $\varepsilon^{\mu\nu\rho}$ we have
 \be
 \begin{split}
  \delta^{(0)}_{\xi}(D_{\nu}\omega_{\rho a}) \ &= \ D_{\nu}(\delta_{\xi}\omega_{\rho a})-\delta^{(0)}_{\xi}A_{\nu}{}^{N}\partial_{N}\omega_{\rho a} \\
  \ &= \ {\cal L}_{\xi}(D_{\nu}\omega_{\rho a})+\xi^{\lambda}[D_{\nu},D_{\lambda}]\omega_{\rho a}+\frac{1}{2}\omega_{\lambda a}[D_{\nu},D_{\rho}]\xi^{\lambda} 
  -\delta^{(0)}_{\xi}A_{\nu}{}^{N}\partial_{N}\omega_{\rho a} 
  \;.
 \end{split}
 \label{Domega}
 \ee 
The commutator is generally given by (\ref{commF}). 
Thus, using this for the $\Lambda$ scalars $\omega$ and $\xi$,  
 \be
     \delta_{\xi}(D_{\nu}\omega_{\rho a}) \ = \ {\cal L}_{\xi}(D_{\nu}\omega_{\rho a})-\xi^{\lambda} F_{\nu\lambda}{}^{N}\partial_N \omega_{\rho a}
     -\frac{1}{2}F_{\nu\rho}{}^{N}\partial_N \xi^{\lambda}\omega_{\lambda a} -\delta^{(0)}_{\xi}A_{\nu}{}^{N}\partial_{N}\omega_{\rho a} 
     \;.
  \ee
As usual for gauge covariant diffeomorphisms, 
the first $F_{\mu\nu}$-term cancels against the same term from $\delta_{\xi}A$, c.f.~(\ref{deltaxiA}). In contrast, 
the second $F_{\mu\nu}$-term survives in the variation, such that
 \be\label{EHvar}
 \begin{split}
  \delta^{(0)}_{\xi}{\cal L}_{\rm EH} \ = \ -2\varepsilon^{\mu\nu\rho} \delta_{\xi}\big(e_{\mu}{}^{a} D_{\nu}\omega_{\rho a}\big) 
  \ = \ \varepsilon^{\mu\nu\rho} e_{\mu}{}^{a} F_{\nu\rho}{}^{N}\partial_{N}\xi^{\lambda} \omega_{\lambda a}
   \;.
 \end{split}
 \ee 
Again, we have suppressed all terms of type ${\cal M}^{MN}\partial_M \otimes \partial_N$ 
induced by the non-covariant transformation of (\ref{deltaxiA}).
For the improved Einstein-Hilbert term, we further 
need the non-covariant variation of $\partial_M e_{\rho a}$, 
which is given by
 \be\label{noncovpartial}
  \Delta^{\rm nc}_{\xi}\big(\partial_M e_{\rho a}\big) \ = \ \partial_M\xi^{\lambda} D_{\lambda} e_{\rho a}-\xi^{\lambda}\partial_M\partial_N A_{\lambda}{}^{N} e_{\rho a}
  + D_{\rho}(\partial_M \xi^{\lambda})\, e_{\lambda a}\;.
 \ee 
Putting everything together, the total variation of the improved Einstein-Hilbert term reads
 \be
 \begin{split}
  \delta^{(0)}_{\xi}\big(e\widehat{R}\big) \ = \ \;&\delta_{\xi}\big(\varepsilon^{\mu\nu\rho} e_{\mu}{}^{a} R_{\nu\rho a}+e e^{a\mu} e^{b\nu} F_{\mu\nu}{}^{M} e_{b}{}^{\rho}
   \partial_M e_{\rho a}\big)\\
  \ = \ \;&\varepsilon^{\mu\nu\rho} e_{\mu}{}^{a} F_{\nu\rho}{}^{N}\partial_{N}\xi^{\lambda} \omega_{\lambda a}
  -ee^{a\mu} e^{b\nu} F_{\mu\nu}{}^{M} e_{b}{}^{\rho} \partial_M\xi^{\lambda} D_{\lambda} e_{\rho a} \\
  &-e e^{a\mu} e^{b\nu} F_{\mu\nu}{}^{M} e_{b}{}^{\rho} D_{\rho}(\partial_M\xi^{\lambda}) e_{\lambda a}\;,
 \end{split}
 \label{totalEH}
 \ee  
 up to total derivatives.
Notice that the second term from (\ref{noncovpartial}) actually drops out in here by the antisymmetry of $F$. 
After some algebra for the spin connection, employing its determined form (\ref{explspinconn}), we find that
 \be
  ee_{b}{}^{\nu}e_{c}{}^{\rho} F_{\nu\rho}{}^M \partial_M\xi^{\lambda} \varepsilon^{abc} \omega_{\lambda a}  
  \ = \ 
  e F_{\nu\rho}{}^M\partial_M\xi^{\lambda}
  \Big(g^{\nu\mu} g^{\rho\sigma}  D_{\mu} g_{\sigma\lambda}-(D_{\lambda} e_{a}{}^{\nu})e^{a\rho}\Big)\;, 
\ee 
such that the variation (\ref{totalEH}) reduces to
\be
 \begin{split}
  \delta^{(0)}_{\xi}\big(e\widehat{R}\big) \ = \ \;&e F^{\mu\nu N} {\cal D}_{\mu}(\partial_N\xi^{\rho} g_{\rho\nu})\;,
 \end{split}
 \label{totalEHA}
 \ee  
up to extra ${\cal O}(A_\mu)$ terms from replacing $D_{\mu}$ by ${\cal D}_{\mu}$.
   
Next, we consider the variation of the Chern-Simons term in (\ref{visionaryAction}), given by
\bea
\delta^{(0)}_\xi {\cal L}_{\rm CS} &=& -\frac1{2\sqrt{2}} \,
\varepsilon^{\mu\nu\rho}\left( (\delta_\xi B_{\mu\,M})\,F_{\nu\rho}{}^M
+(\delta_\xi A_{\mu}{}^{M})\,G_{\nu\rho}{}_M\right)
\nonumber\\[1ex]
&=&
-\frac1{2\sqrt{2}} \,
\varepsilon^{\mu\nu\rho}\Big(
{\cal M}^{MN}g_{\mu\lambda}\partial_{N}\xi^{\lambda}\,G_{\nu\rho}{}_M
+f_{MN}{}^K\,F_{\mu\lambda}{}^N\partial_K\xi^\lambda\,F_{\nu\rho}{}^M
\nonumber\\
&&{}\qquad\qquad
+\frac12\,
({\cal M}_{QN}\partial_M{\cal M}^{PN})\, f^{QK}{}_P\partial_K\xi^\lambda g_{\lambda\mu} \,F_{\nu\rho}{}^M
\nonumber\\[1ex]
&&{}\qquad\qquad
+f_{MNK}\partial^N
({\cal M}^{KL}g_{\mu\lambda}\partial_{L}\xi^{\lambda})\,F_{\nu\rho}{}^M
\Big)
-e F^{\mu\nu N} {\cal D}_{\mu}(\partial_N\xi^{\rho} g_{\rho\nu})
\;,
\label{D_CS}
\eea
and we recognize a first conspiracy between the last term and (\ref{totalEHA}).

The kinetic term in (\ref{visionaryAction}) is not invariant under the full diffeomorphisms either,
due to anomalous terms of similar origin as in~(\ref{Domega}). 
Specifially, with (\ref{deltaxiA}), (\ref{varB}), the scalar current transforms as
\bea
  \delta^{(0)}_{\xi}\big({\cal D}_{\mu}{\cal M}_{MN}\big) &=& 
  {\cal L}_{\xi}\big(\cD_{\mu}{\cal M}_{MN}\big)+4F_{\mu\nu}{}^{P} {\cal M}_{P(M} \partial_{N)}\xi^{\nu}
  -4 F_{\mu\nu (M} {\cal M}_{N)}{}^{P}\partial_P \xi^{\nu}
  \nonumber\\
  &&{}\hspace{-1.5cm}
  -{\cal M}^{KQ}  (\partial_K\,{\cal M}_{MN}) \partial_{Q}\xi^{\nu}g_{\mu\nu}
  -4{\cal M}_{K(M}\partial_{N)} ({\cal M}^{KQ}\partial_{Q}\xi^{\nu}g_{\mu\nu}) 
  \nonumber\\
&&{}\hspace{-1.5cm}
  +4\partial^K (\partial^{Q}\xi^{\nu}g_{\mu\nu} {\cal M}_{Q(M}) {\cal M}_{N)K}
  {}+({\cal M}_{QS}\partial_K{\cal M}^{PS}) f^{QR}{}_P\,\partial_R\xi^\lambda g_{\lambda\mu}
  \,f_{K(M}{}^L{\cal M}_{N)L}
  \nonumber\\
&&{}\hspace{-1.5cm}
  -2\,f_{K(M}{}^L\,{\cal M}_{N)L}\, g_{\mu\lambda} \, e^{-1}\varepsilon^{\lambda\nu\rho}
\,{\cal D}_\nu\left(\partial_K \xi^\sigma g_{\rho\sigma} \right)
\;.
\eea
Up to total derivatives, the kinetic term thus varies into
\bea
\delta^{(0)}_\xi {\cal L}_{\rm kin} &=& 
  e\,{\cal D}^\mu {\cal M}^{MN}\, F_{\mu\nu}{}^{P} {\cal M}_{PM} \partial_{N}\xi^{\nu}  
  \nonumber\\
  &&{}
  -e\,{\cal D}_\mu {\cal M}^{MN}\,\Big(
  {\cal M}_{KM}\,\partial_{N} {\cal M}^{KQ}
  +\frac18\,{\cal M}^{KQ}  \,\partial_K {\cal M}_{MN}
  -\frac12\, \partial_M {\cal M}_{N}{}^{Q} \Big)\, \partial_{Q}\xi^\mu  
  \nonumber\\
&&{}
  -\frac1{2\sqrt{2}} \, \varepsilon^{\mu\nu\rho}\,f_{KM}{}^L{\cal M}_{NL}\,
  {\cal D}_\mu {\cal M}^{MN}
\,{\cal D}_\nu\left(\partial_K \xi^\sigma g_{\rho\sigma} \right)
 \;.
\label{Dkin}
\eea
After partial integration, the last term takes the form
\bea
&&
-\frac1{4\sqrt{2}}\, \varepsilon^{\mu\nu\rho} \,
[{\cal D}_{\mu},{\cal D}_\nu]{\cal M}^{MN}\,f_{KM}{}^P\,{\cal M}_{PN}\,\partial^K \xi^\lambda\,g_{\lambda\rho}
\nonumber\\
&&{}
+\frac1{2\sqrt{2}}\,\varepsilon^{\mu\nu\rho} \,
{\cal D}_\mu{\cal M}^{MN}{\cal D}_\nu{\cal M}_{PN}\,f_{KM}{}^P\,\partial^K \xi^\lambda\,g_{\lambda\rho}
\;,
\label{x00}
\eea
of which the commutator term reduces to
\bea
&&{}
\frac1{2\sqrt{2}}\, \varepsilon^{\mu\nu\rho} {\cal M}^{MN}\,
G_{\mu\nu M}\partial_N\xi^\lambda g_{\lambda\rho}
\label{0varGF}
\\
&&{}
+\frac1{2\sqrt{2}}\,\varepsilon^{\mu\nu\rho}\,F_{\mu\nu}{}^M
\,\Big(
\frac12({\cal M}_{QN}\partial_M{\cal M}^{PN}) f^{QK}{}_P\partial_K\xi^\lambda g_{\lambda\rho}
+f_{MNK}\partial^N({\cal M}^{KL}\partial_L\xi^\lambda g_{\lambda\rho})
\Big)
\;,
\nonumber
\eea
up to total derivatives,
and entirely cancels against the corresponding terms in the variation (\ref{D_CS}) of the Chern-Simons term.

Finally, we consider the anomalous variation of the potential $V({\cal M},g)$. This is due to the transformations
 \bea
 \label{diffdM}
  \delta_{\xi}(\partial_K {\cal M}_{MN}) &=& \xi^{\mu}{\cal  D}_{\mu}(\partial_K{\cal M}_{MN})
  +\partial_K\xi^{\mu} \,{\cal  D}_{\mu}{\cal M}_{MN}
  -4\partial_K(\partial_{(M} A_{\mu}{}^{P}-\partial^{P} A_{\mu (M}){\cal M}_{N)P}    
  \nonumber\\
  &&{}+\xi^\mu (\partial_L A_\mu{}^L) (\partial_K {\cal M}_{MN})
  +2\,\xi^\mu\partial_K \tilde{B}_\mu{}^L\,f_{L(M}{}^P {\cal M}_{N)P}
  \;,\nonumber\\[1ex]
\delta_{\xi} (\partial_M g_{\mu\nu}) &=&
{\cal L}_\xi (\partial_M g_{\mu\nu}) + (\partial_M\xi^\rho) D_\rho g_{\mu\nu}+
2g_{\rho(\mu} {\cal  D}_{\nu)}(\partial_M \xi^\rho)
\nonumber\\
&&{}
-2\xi^\rho g_{\mu\nu}\,(\partial_M\partial_N A_\rho{}^N)
+\xi^\rho(\partial_N A_\rho{}^N)\,(\partial_M g_{\mu\nu})
+2(\partial_N A_{(\mu}{}^N)\,(\partial_M\xi^\rho)\,g_{\nu)\rho}
\;,
\nonumber\\[1ex]
\delta_{\xi} (\partial_M g) &=&
{\cal L}_\xi (\partial_M g) +(\partial_M\xi^\mu)D_\mu g+2g {\cal  D}_\mu(\partial_M\xi^\mu)
-6 \,\xi^\mu\,g\, \partial_M\partial_N A_\mu{}^N 
\nonumber\\
&&{}
+ \xi^\mu (\partial_N A_\mu{}^N)\,\partial_M g
+2g(\partial_N A_\mu{}^N)\,\partial_M\xi^\mu
\;.
\eea
Again, we suppress in the following all terms with explicit appearance of the gauge fields,
as these cancel separately.
The variation of the potential then gives rise to
\bea
\delta_\xi (-eV({\cal M},g))&=& 
\frac{3}{8}\,e\, \partial_L\xi^{\mu} \,D_{\mu}{\cal M}_{MN} \, \partial_K {\cal M}^{MN}  {\cal M}^{KL}
-\frac32\,e\,\partial_L \xi^\mu\,D_\mu{\cal M}_{MN} \, \partial^M {\cal M}^N{}_{K} {\cal M}^{KL}
\;.
\nonumber\\
\label{varpot}
\eea

Collecting all the terms that we have encountered in 
(\ref{totalEHA}), (\ref{D_CS}), (\ref{Dkin})--(\ref{0varGF}), and (\ref{varpot}),
we are left with
\bea
\delta^{(0)} {\cal L} &=&
 \frac14\,e\,{\cal D}_\mu {\cal M}^{MN}\,\Big(
  \partial_K {\cal M}_{MN}
  -2\,\partial_{N}{\cal M}_{KM}
  +2\, {\cal M}_{KL}\,\partial_M {\cal M}_{N}{}^{L}
   \Big)\,  {\cal M}^{KQ}\,\partial_{Q}\xi^\mu  
\nonumber\\
&&{}
-\frac1{2\sqrt{2}} \,
\varepsilon^{\mu\nu\rho}\,f_{MN}{}^K\partial_K\xi^\lambda
\left( F_{\mu\nu}{}^M\,F_{\rho\lambda}{}^N
-
{\cal D}_\mu{\cal M}^{ML}\,{\cal D}_\nu{\cal M}_L{}^N\,g_{\lambda\rho} \right)
  \nonumber\\
  &&{}
+e\,{\cal D}^\mu {\cal M}^{MN}\, F_{\mu\nu}{}^{P} {\cal M}_{PM} \partial_{N}\xi^{\nu}  
\;.
\eea
Some algebra (e.g., using an explicit parametrization as in (\ref{VV}) below) 
shows that the first line in this expression is actually vanishing,
while the remaining terms can be recast into the compact form
\bea
\delta^{(0)}_\xi S  &=&  -\frac1{2\sqrt{2}}\,  \int d^3x \, d^3Y 
\varepsilon^{\mu\nu\rho} \, f_{MN}{}^K \,\partial_{K}\xi^{\lambda}\,
{\cal E}^{(A)}_{\mu\nu\vphantom{M}}{}^{M}
{\cal E}^{(A)}_{\rho\lambda\vphantom{M}}{}^{N}
\;,
\label{obstruct}
\eea
in terms of the duality equation (\ref{varLAB}). I.e.\ the anomalous variation of the action comes out to be proportional to the duality equations
(\ref{varLAB}) obtained by varying the Lagrangian w.r.t.\ to the vector fields. We conclude that invariance
of the action can be achieved by properly modifying the transformation law (\ref{varB}) for the vector field $B_{\mu\,M}$.
However, naively modifying the transformation law (\ref{varB}) induces a variation of the vector field, that is no longer
compatible with the constraints (\ref{fullconstraints}), rendering the diffeomorphism symmetry inconsistent.
Fortunately, this can be remedied by modifying the transformation laws (\ref{deltaxiA}), (\ref{varB}) by another
trivial (formal `equations-of-motion'-)symmetry. 
The resulting full diffeomorphism transformations of the vector fields take the form
\bea
\delta_\xi A_{\mu}{}^{M} &\equiv&  
\delta^{(0)}_\xi A_{\mu}{}^{M} -\xi^\nu\,{\cal E}^{(A)}_{\nu\mu\vphantom{M}}{}^M
\;,
\nonumber\\
\delta_\xi B_{\mu\,M} &\equiv& 
\delta^{(0)}_\xi B_{\mu\,M} + f_{KM}{}^N \,\partial_{N}\xi^{\nu}\,
{\cal E}^{(A)}_{\mu\nu\vphantom{M}}{}^{K} -\xi^\nu\,{\cal E}^{(B)}_{\nu\mu\,M}
\;.
\label{varAB}
\eea
The second term in the variation of $B_{\mu\,M}$ is precisely necessary in order to cancel the anomalous variation of (\ref{obstruct}), such that the action becomes invariant.
The last terms in the variation of $A_\mu{}^M$ and $B_{\mu\,M}$, respectively, constitute an `equations-of-motion'-symmetry
of the Lagrangian, as follows immediately from (\ref{varLAB}) and (\ref{varAB}), and thus do not corrupt the invariance of the action.
Their presence however is crucial in order to maintain compatibility of the transformation laws (\ref{varAB}) with
the constraints (\ref{fullconstraints}) imposed on the combination of vector fields ${\tilde{B}}_{\mu\,M}$. To show this, we calculate
from (\ref{varAB})
\bea
\delta_\xi \left(B_{\mu\,M}-f_{KM}{}^{N}\partial_N A_\mu{}^{K} \right) &=& 
\xi^\nu\left(
G_{\nu\mu\,M}-f_{KM}{}^{N}\partial_N F_{\nu\mu}{}^{K}
\right)
- \xi^\nu\left({\cal E}^{(B)}_{\nu\mu\,M}
- f_{KM}{}^{N}\partial_N  {\cal E}^{(A)}_{\nu\mu\vphantom{M}}{}^K\right)
\nonumber\\
&&{}
+{\cal O}_{\mu\,M}
\;,
\label{compB}
\eea
where by ${\cal O}_{\mu\,M}$ we collect all terms that are separately compatible with the constraints
(\ref{fullconstraints}), i.e.
\bea 
{\mathbb P}_{KL}{}^{MN}\,{\cal O}_{\mu\,M}\otimes C_N   \ = \ 0\;, \qquad \forall \; C_M \ = \ (\partial_M , \tilde{B}_M,\tilde{\Sigma}_M)\;.
\label{compO}
\eea
Specifically, in (\ref{compB}) all terms collected in ${\cal O}_{\mu\,M}$ carry an explicit derivative $\partial_M$,
such that (\ref{compO}) is manifest. Using the explicit form of (\ref{duality}), a quick calculation shows that
the first two terms on the r.h.s.\ of (\ref{compB}) mutually cancel, leaving only the ${\cal O}_{\mu M}$ term. 
Thus we conclude that
\bea 
{\mathbb P}_{KL}{}^{MN}\,\delta_\xi \tilde{B}_{\mu\,M}\otimes C_N   \ = \ 0\;, 
\qquad \forall \; C_M \ = \ (\partial_M , \tilde{B}_M,\tilde{\Sigma}_M)\;,
\label{compBtO}
\eea
as required for consisteny.

Summarizing, we have shown that the action~(\ref{visionaryAction})
is invariant under generalized diffeomorphisms with parameters $\xi^\mu(x,Y)$,
provided the fields transform as (\ref{xidiff}), (\ref{varAB}).
It is remarkable, that in the final transformation law for the vector fields,
all terms carrying explicit field strengths drop out, e.g.\
\bea
\delta_\xi A_{\mu}{}^{M} &=&  
 {\cal M}^{MN} g_{\mu\nu}\partial_{N}\xi^{\nu}
-\frac1{2\sqrt{2}} \,e\varepsilon_{\mu\nu\rho} \, \xi^\nu f^{MK}{}_{L}\,
{\cal D}^\rho {\cal M}^{LN} {\cal M}_{NK} 
\;.
\label{diffA}
\eea
This reflects the fact that the vector fields in this three-dimensional formulation do
not carry propagating degrees of freedom, but are related to the scalar fields
by means of the duality equations~(\ref{duality}).
Indeed, similar structures arise in three-dimensional supergravity, where the supersymmetry
algebra closes into transformations of the type (\ref{diffA}) rather than into the standard
covariant form~\cite{deWit:2008ta}.

\section{Reduction to $D=4$ Einstein gravity}
In this section we verify that by explicitly solving the constraint (\ref{fullconstraints}) and
reducing to fields that depend only on four coordinates, we recover precisely $D=4$ Einstein gravity. 
To this end we rewrite in the first subsection Einstein gravity \`a la Kaluza-Klein via a $(3+1)$
splitting of the coordinates, reviewing the results of \cite{Hohm:2005sc,Hohm:2006ud}. 
We stress that this does not involve any truncation, as we keep 
the dependence on all four coordinates. In the second subsection we show that the 
$(3+3)$-dimensional theory reduces to Einstein gravity in this formulation.

\subsection{$(3+1)$ splitting of $D=4$ Einstein gravity}

In this section, we recast four-dimensional Einstein gravity into the form of a three-dimensional
gravitational theory by rearranging the fields in Kaluza-Klein form but keeping the full
dependence on the fourth coordinate $y$.
We follow~\cite{Hohm:2005sc,Hohm:2006ud}, see also~\cite{Aulakh:1985un,Cho:1992rq}.
We start from the $D=4$ Einstein-Hilbert action (with mostly plus signature), 
 \be
  S_{\rm EH} \ = \ \int d^4x\,e\,R \ \equiv \  \int d^4x\,e\, e_{\hat{a}}{}^{\hat{\mu}}e_{\hat{b}}{}^{\hat{\nu}} R_{\hat{\mu}\hat{\nu}}{}^{\hat{a}\hat{b}}\;, 
  \label{S4D}
 \ee 
where $D=4$ world and Lorentz indices are denoted 
by $\hat{\mu}, \hat{\nu},\ldots$ and $\hat{a}, \hat{b},\ldots$, respectively. 
Next, we perform a splitting of coordinates, $x^{\hat{\mu}}=(x^{\mu},y)$, and indices, 
$\hat{\mu}=(\mu,3)$, etc., and 
reduce the Lorentz gauge symmetry to $SO(1,2)$ by choosing an upper-triangular gauge, 
 \be\label{KKbein}
  e_{\hat{\mu}}{}^{\hat{a}} \ = \  \left(\begin{array}{cc} \phi^{-1/2}e_{\mu}{}^a &
  \phi^{1/2} A_{\mu} \\ 0 & \phi^{1/2} \end{array}\right)\;.
 \ee 
In the following it will be convenient to have the action of the full four-dimensional 
diffeomorphisms (parameterized by $\xi^{\hat{\mu}}=(\xi^{\mu},\Lambda$)) at our disposal. Applying 
 \be
  \delta_{\xi}e_{\hat{\mu}}{}^{\hat{a}} \ = \ \xi^{\hat{\nu}}\partial_{\hat{\nu}}e_{\hat{\mu}}{}^{\hat{a}} 
  +\partial_{\hat{\mu}}\xi^{\hat{\nu}} e_{\hat{\nu}}{}^{\hat{a}}\;,
 \ee  
to (\ref{KKbein}) we have to add a compensating local Lorentz transformation, 
with parameter $\lambda^a{}_{3}=\phi^{-1}\partial_y\xi^{\nu} e_{\nu}{}^{a}$, in order to 
preserve the gauge choice. 
Under $\xi^{\mu}$ diffeomorphisms, the fields thus transform as
  \begin{equation}\label{diff}
  \begin{split}
    \delta_{\xi}e_{\mu}{}^a \ & = \  \xi^{\rho}\partial_{\rho}e_{\mu}{}^a+\partial_{\mu}\xi^{\rho}e_{\rho}{}^a
   +\partial_y \xi^{\rho}A_{\rho}e_{\mu}{}^a\;, \\
   \delta_{\xi}A_{\mu} \ &= \  \xi^{\rho}\partial_{\rho}A_{\mu}
   + \partial_{\mu}\xi^{\rho}A_{\rho}-A_{\mu}\partial_y\xi^{\rho}A_{\rho}+\phi^{-2}\partial_y\xi^{\nu} g_{\nu\mu}\;,\\
   \delta_{\xi}\phi \  & = \  \xi^{\rho}\partial_{\rho}\phi
   +2\phi\,\partial_y\xi^\rho A_\rho
 \;.
  \end{split}
 \end{equation}
Under dimensional reduction, i.e.\ $\partial_y = 0$ this reduces to the standard $D=3$
diffeomorphism transformations. 
 The action of the four-dimensional $\Lambda$ diffeomorphisms takes the form
  \be\label{LambdaDiff}
 \begin{split}
   \delta_{\Lambda}e_{\mu}{}^{a} \ &= \ \Lambda\partial_y e_{\mu}{}^{a}+\partial_y\Lambda e_{\mu}{}^{a}\;, \\
  \delta_{\Lambda}A_{\mu} \ &= \ \partial_{\mu}\Lambda+\Lambda\partial_y A_{\mu}-A_{\mu}\partial_y\Lambda\;, \\
  \delta_{\Lambda}\phi \ &= \ \Lambda \partial_y\phi+2\phi \partial_y\Lambda\;,   
 \end{split}
 \ee  
of an infinite-dimensional non-abelian gauge structure in three dimensions.
Accordingly, we can define covariant derivatives and field strengths for 
the $\Lambda$ transformations, as
 \begin{equation}
  \begin{split}
   D_{\mu}e_{\nu}{}^{a} \ &= \ \partial_{\mu}e_{\nu}{}^{a}-A_{\mu}\partial_y e_{\nu}{}^{a}-e_{\nu}{}^{a}\partial_y A_{\mu}\;, \\
   F_{\mu\nu} \ &= \  \partial_{\mu}A_{\nu}-\partial_{\nu}A_{\mu}-A_{\mu}
   \partial_yA_{\nu}+A_{\nu}\partial_yA_{\mu}\;, \\
   D_{\mu}\phi \ &= \  \partial_{\mu}\phi-A_{\mu}\partial_y\phi
   -2\phi\partial_yA_{\mu}\;.
  \end{split}
  \label{dyo}
 \end{equation}

The complete action (\ref{S4D}) can then be expressed
in terms of manifestly $\Lambda$-covariant objects as
 \bea
   S_{\text{EH}}=\int d^3 x\,dy\, e\left[R^{(3),\text{cov}}
   -\frac{1}{4}\phi^2F^{\mu\nu}F_{\mu\nu}-\frac{1}{2}\phi^{-2}g^{\mu\nu}
   D_{\mu}\phi D_{\nu}\phi - {\cal L}_m\right]\;,
 \eea
where $R^{(3),\text{cov}}$ denotes the generalized Ricci scalar 
with respect to the covariantized connection (\ref{dyo}).
The last term reads
 \bea\label{spin2mass}
   {\cal L}_m= \frac{1}{4}\phi^{-2}g^{\mu\nu}
   g^{\rho\sigma}(D_yg_{\mu\rho}D_yg_{\nu\sigma}-D_yg_{\mu\nu}
   D_yg_{\rho\sigma})
   +e^{a\mu}e^{b\nu}F_{\mu\nu}\,e_b{}^{\rho}\partial_y e_{\rho a}\;.
 \eea
with $D_y g_{\mu\nu} \equiv\partial_y g_{\mu\nu} - g_{\mu\nu}\, \phi^{-1}\partial_y\phi$\,.
Upon some rearrangement, the $D=4$ Lagrangian takes the form
 \be\label{4DLagr}
 \begin{split}
  {\cal L} \ = \ e\Big(&\widehat{R}-\frac{1}{4}\,\phi^2 F^{\mu\nu} F_{\mu\nu}
  -\frac{1}{2}\,\phi^{-2} g^{\mu\nu} D_{\mu}\phi\,D_{\nu}\phi \\  
 &+\frac{3}{2}\,\phi^{-4}(\partial_y\phi)^2-g^{-1}\partial_yg\,\phi^{-3}\partial_y\phi
 +\frac{1}{4}\,\phi^{-2}(g^{-1}\partial_yg)^2
  +\frac{1}{4}\,\phi^{-2}\partial_y g^{\mu\nu}\,\partial_y g_{\mu\nu}\Big)\;, 
\end{split}
\ee 
with the `improved' Einstein-Hilbert term given by the  Lorentz invariant combination
 \be
  e \widehat{R} \ = \ - \varepsilon^{\mu\nu\rho} e_{\mu}{}^{a} R^{(3),\text{cov}}_{\nu\rho a} - ee^{a\mu}e^{b\nu} F_{\mu\nu} e_{b}{}^{\rho}\partial_y e_{\rho a }\;.
  \label{impEH}
 \ee

\subsection{$(3+1)$ Chern-Simons form of $D=4$ Einstein gravity}

We have rewritten four-dimensional Einstein gravity in the form~(\ref{4DLagr})
reminiscent of the three-dimensional Kaluza-Klein form. Indeed, upon dimensional
reduction $\partial_y=0$, this action reduces to the standard form of a Maxwell and
a scalar field coupled to three-dimensional gravity. In that case, the three-dimensional
duality symmetry $SL(2,\mathbb{R})$ is made manifest by dualizing the Maxwell field
into another scalar giving rise to a $SL(2,\mathbb{R})/SO(2)$ target space~\cite{Ehlers:1957}.
A similar construction is possible for the full four-dimensional theory~\cite{Hohm:2005sc}.
Since due to the $y$-dependence the modes of the Kaluza-Klein vector carry a non-abelian 
gauge structure, their dualization necessitates the introduction of additional 
non-propagating vector fields~\cite{Nicolai:2003bp}. The resulting theory takes the
form of a scalar sigma-model coupled to Chern-Simons vectors. 

To this end, we introduce the dual scalar $\varphi$ by means of the duality equation
together with a vector field $B_\mu$
\bea
{\cal D}_{\mu}\varphi
&\equiv& \partial_{\mu}\varphi-A_{\mu}\partial_y\varphi
   -2\varphi\partial_yA_{\mu}+B_\mu
~\equiv~\frac{1}{2}\,e\varepsilon_{\mu\nu\rho} \,\phi^{2} F^{\nu\rho}
\;,
\label{varphi}
\eea
such that the Bianchi identity and the Yang-Mills field equation for $F_{\mu\nu}$ 
give rise to the field and duality equations for $\varphi$ and $B_\mu$, respectively.
On the level of the action~(\ref{4DLagr}), this corresponds to replacing
\bea
-\frac{1}{4}\,\phi^2 F^{\mu\nu} F_{\mu\nu} &\longrightarrow&  
-\frac{1}{2}\,\phi^{-2}g^{\mu\nu}{\cal D}_{\mu}\varphi\,{\cal D}_{\nu}\varphi
+\frac{1}{2}\,\varepsilon^{\mu\nu\rho} B_{\mu} F_{\nu\rho}\;.
\label{YMCS}
\eea
The Yang-Mills form of the action is then recovered by integrating out $B$. 
The Lagrangian of four-dimensional Einstein gravity in this formulation thus
is given by
  \be\label{4DLagrCS}
 \begin{split}
  {\cal L} \ = \ e\Big(&\widehat{R}
  -\frac{1}{2}\,\phi^{-2} g^{\mu\nu} \left(D_{\mu}\phi\,D_{\nu}\phi 
  +{\cal D}_{\mu}\varphi\,{\cal D}_{\nu}\varphi\right)
+\frac{1}{2}\varepsilon^{\mu\nu\rho} B_{\mu} F_{\nu\rho}\\  
 &+\frac{3}{2}\,\phi^{-4}(\partial_y\phi)^2-g^{-1}\partial_yg\,\phi^{-3}\partial_y\phi
 +\frac{1}{4}\,\phi^{-2}(g^{-1}\partial_yg)^2
  +\frac{1}{4}\,\phi^{-2}\partial_y g^{\mu\nu}\,\partial_y g_{\mu\nu}\Big)
  \;,
\end{split}
\ee 
with the `improved' Ricci scalar from (\ref{impEH}).
Under dimensional reduction $\partial_y=0$, it reduces to the form
of the three-dimensional theory in which (modulo integrating out the vector fields)
the global duality symmetry $SL(2,\mathbb{R})$ is manifest.

We deduce from (\ref{varphi}), 
that the action of $\Lambda$-transformations~(\ref{LambdaDiff}) on the new fields
is given by
\be
   \delta_{\Lambda}\varphi ~=~ \Lambda \partial_y\varphi+2\varphi \partial_y\Lambda\;, 
   \qquad
   \delta_{\Lambda} B_\mu ~=~ \Lambda \partial_y B_\mu+2B_\mu \partial_y\Lambda\;.
\ee  
It is slightly more involved to derive the transformation law for $B_\mu$
under covariantized diffeomorphisms with the remaining fields transforming as
\bea
  \delta_{\xi} \phi &=& \xi^\mu D_\mu \phi\;,\qquad  \delta_{\xi} \varphi ~=~ \xi^\mu {\cal D}_\mu\,\varphi\;,\nonumber\\
  \delta_{\xi} A_{\mu} &=&  \xi^{\nu} F_{\nu\mu}+\phi^{-2}\partial_y\xi^{\nu} g_{\nu\mu}
  ~\equiv~  \xi^{\nu} F_{\nu\mu}+\Delta^{\rm nc}_\xi A_\mu
  \;,
  \label{varA4}
 \eea
under a proper combination of (\ref{diff}) and (\ref{LambdaDiff}). 
The transformation of $B_\mu$ then is fixed from requiring that the total
variation of the action remains unchanged under the replacement (\ref{YMCS}).
With the variation of the Yang-Mills term given by
\be\label{YMvar}
  \delta_{\xi}\Big(-e\frac{1}{4}\phi^2 F^{\mu\nu} F_{\mu\nu}\Big) \ = \ 
 -e\phi^2 F^{\mu\nu} D_{\mu}\big(\Delta^{\rm nc}_\xi A_\nu \big)  
\;,
 \ee
it is straightforward to derive that
invariance of the action under the replacement (\ref{YMCS})
requires that
\bea
\delta_\xi\left({\cal D}_\mu\varphi\right) &=& {\cal L}_\xi\left( {\cal D}_\mu\varphi \right)+
e^{-1}g_{\mu\nu}\,\varepsilon^{\nu\lambda\rho}\phi^2D_{\lambda}\big(\phi^{-2}\partial_y \xi^{\sigma} g_{\sigma \rho}\big) 
\;.
\eea
With the explicit covariant derivative defined in (\ref{varphi}), we deduce that
 \be
  \delta_{\xi}B_{\mu} \ = \ \xi^{\nu}G_{\nu\mu}-2\varphi F_{\mu\nu}\partial_y\xi^{\nu}
  +\Delta_{\xi}A^{\rm nc}_{\mu}\,\partial_y\varphi
  +2\varphi\, \partial_y(\Delta^{\rm nc}_{\xi}A_{\mu})
  +e^{-1}g_{\mu\nu}\,\varepsilon^{\nu\lambda\rho}\phi^2D_{\lambda}\big(\Delta^{\rm nc}_\xi A_{\rho}\big) 
  \;,
  \label{varB4}
\ee 
under covariant diffeomorphisms, 
where we defined the field strength of $B_{\mu}$, 
 \be
  G_{\mu\nu} \ = \ D_{\mu}B_{\nu}-D_{\nu}B_{\mu}\;, 
 \ee
using $\Lambda$ covariant derivatives.   
Under dimensional reduction $\partial_y=0$, this reduces to the standard
transformation law of three-dimensional vector fields.

\subsection{Reduction of the $(3+3)$-dimensional theory}

In this section, we consider the $(3+3)$-dimensional action~(\ref{visionaryAction})
and show that after explicitly solving the section condition (\ref{fullconstraints})
this action reduces to (\ref{4DLagrCS}) which we have obtained as an equivalent
reformulation of four-dimensional Einstein gravity.

To start with, we choose an explicit parametrisation of the matrix ${\cal V}$
from (\ref{matrixV}) in triangular gauge. Denoting the basis of the Lie algebra 
$\frak{sl}(2,\mathbb{R})$ by $\{e,h,f\}$ we write 
 \be
  {\cal V} \ = \ \exp(\sqrt{2}\varphi f)\exp(-\ln\phi \,h)
  \ = \ \begin{pmatrix}    \phi &\sqrt{2} \varphi & -\varphi^2 \phi^{-1} \\[0.5ex]
  0 & 1 & -\sqrt{2}\varphi \phi^{-1} \\[0.5ex] 0 & 0 & \phi^{-1} \end{pmatrix}\;.
 \label{VV}
 \ee 
 In this basis, we normalise the antisymmetric
 structure constants $f_{MNK}$ by $f_{hef}=1$, and take the metric $\eta_{MN}$ of the form (\ref{eta}),
 i.e.\ $\eta_{hh}=\eta_{ef}=1$\,.
Next, we choose an explicit solution to the section constraints 
(\ref{weakerstrong}), (\ref{stronger}) by restricting the $Y^M$ dependence of all fields to
a single coordinate $y\equiv Y^e$, such that derivatives $\partial_M$ reduce to
\bea
\partial_h &=& \partial_f ~=~ 0
\;.
\label{secder}
\eea 
Similarly, we solve (\ref{fullconstraints}) by setting $\tilde B_{\mu}{}^{h} = \tilde B_{\mu}{}^e = 0$, 
implying that
\bea
B_{\mu}{}^{h} + \partial_y A_\mu{}^e &=& 0 ~=~ B_{\mu}{}^e\;,
\label{secB}
\eea
for the components of the vector field $B_{\mu}{}^{M}$\,. With this choice, 
a short calculation reveals that the kinetic and the Chern-Simons term of 
the action (\ref{visionaryAction}) reduce to
\bea
\frac1{16}{\cal D}_\mu{\cal M}_{MN}\,{\cal D}^\mu{\cal M}^{MN} \Big|_{(\ref{VV})-(\ref{secB})}
&=&
-\frac{1}{2\phi^{2}} \left(D_\mu \phi D^\mu \phi
+{\cal D}_\mu \varphi {\cal D}^\mu \varphi\right)\;,
\nonumber\\
-\frac1{2\sqrt{2}}\,\varepsilon^{\mu\nu\rho} \,
B_{\mu\,M} F_{\nu\rho}{}^M \Big|_{(\ref{VV})-(\ref{secB})}
&=&
\frac1{2}\,\varepsilon^{\mu\nu\rho} \,
{B}_{\mu} \,F_{\nu\rho}
\;,
\eea
reproducing the corresponding terms in the Lagrangian (\ref{4DLagrCS}), 
with covariant derivatives and field strength from (\ref{dyo}),
upon the identification
\bea
A_\mu ~\equiv~ A_\mu{}^e\;,\qquad B_\mu ~\equiv~
-\frac1{\sqrt{2}} \left(B_\mu{}^f+\partial_y A_\mu{}^h\right)
\;.
\label{AB}
\eea
In particular, with this solution of the section constraint, the Lagrangian depends only on the
two remaining combinations (\ref{AB}) of the original vector fields $A_\mu{}^M$ and ${B}_{\mu\,M}$,
which take the role of the vector fields of the Lagrangian (\ref{4DLagrCS}).

It remains to calculate the form of the scalar potential (\ref{scalar}) in this parametrisation and
 after plugging in (\ref{secder}). A straightforward calculation confirms that
 \bea
 V({\cal M},g) \Big|_{(\ref{VV})-(\ref{secB})} &=&
 -\frac{3}{2}\,\phi^{-4}(\partial_y\phi)^2
 +g^{-1}\partial_yg\,\phi^{-3}\partial_y\phi
 \nonumber\\
 &&{}
 -\frac{1}{4}\,\phi^{-2}\left((g^{-1}\partial_yg)^2
 +\partial_y g^{\mu\nu}\,\partial_y g_{\mu\nu}\right)
  \;,
 \eea
 which is in agreement with (\ref{4DLagrCS}). 
In particular, there is no dependence left on $\varphi$. 
 This finishes our demonstration that the action 
 (\ref{visionaryAction}) reduces to four-dimensional Einstein gravity
(in the form (\ref{4DLagrCS})) upon explicitly solving the section condition (\ref{fullconstraints}). 

To complete this section, it is instructive to consider the action induced by the $(3+3)$-dimensional
diffeomorphisms~(\ref{varAB}) on the vector fields that survive in the $(3+1)$-dimensional action. 
Evaluating these transformations for the specific vector components~(\ref{AB}) upon 
imposing (\ref{secder}), (\ref{secB}),
leads to
\bea
\delta_\xi A_{\mu} &=&  
 \phi^{-2}\left(
 g_{\mu\nu}\partial_{y}\xi^{\nu}
+e\varepsilon_{\mu\nu\rho} \, \xi^\nu {\cal D}^\rho \varphi \right)
\;,
\nonumber\\
\delta_{\xi} B_{\mu} 
&=&
\partial_y\left( 2 \phi^{-2}\varphi \, g_{\mu\nu}\,\partial_y\xi^\nu\right)
-\phi^{-2}\,\partial_y\varphi\,g_{\mu\nu}\,\partial_y\xi^\nu
+ g_{\mu\lambda} \, e^{-1}\varepsilon^{\lambda\nu\rho}
\,{\cal D}_\nu\left(\partial_y \xi^\sigma g_{\rho\sigma} \right)
\nonumber\\
&&{}
+\varepsilon_{\mu\nu\rho}\,\Big(
2\partial_y \left(e\xi^\nu\,\phi^{-2}
\left(\phi D^\rho \phi +\varphi {\cal D}^\rho \varphi \right)\right)
-\phi^{-2} \,e\xi^\nu\,\left(
\partial_y \phi D^\rho \phi +\partial_y \varphi {\cal D}^\rho \varphi \right)
\Big) 
\nonumber\\
&&{}
+\frac1{\sqrt{2}}\, \xi^\nu\,\Omega_{\nu\mu\,e}
\;.
\eea
Modifying these transformation laws by a standard
equations-of-motion-symmetry 
\bea
\delta A_\mu \rightarrow \delta A_\mu 
+ e\varepsilon_{\mu\nu\rho}\,\xi^\nu\frac{\partial {\cal L}}{\partial B_\rho}
\;,\qquad
\delta B_\mu \rightarrow \delta B_\mu 
+ e\varepsilon_{\mu\nu\rho}\,\xi^\nu\frac{\partial {\cal L}}{\partial A_\rho}
\;,
\label{eoms}
\eea
which separately leaves the action (\ref{4DLagrCS}) invariant,
the transformation laws take the more familiar form
\bea
\delta^{\rm mod}_\xi A_{\mu} &=&  
\xi^\nu F_{\nu\mu}+ \phi^{-2}g_{\mu\nu}\partial_{y}\xi^{\nu}
\;,
\nonumber\\
\delta^{\rm mod}_{\xi} B_{\mu} 
&=&
\xi^\nu G_{\nu\mu}+
\varphi \,\partial_y\left( 2 \phi^{-2} g_{\mu\nu}\,\partial_y\xi^\nu\right)
+\phi^{-2}\,\partial_y\varphi\,g_{\mu\nu}\,\partial_y\xi^\nu
+ g_{\mu\lambda} \, e^{-1}\varepsilon^{\lambda\nu\rho}
\,{\cal D}_\nu\left(\partial_y \xi^\sigma g_{\rho\sigma} \right)
\nonumber\\
&&{}
+2e\,\varepsilon_{\mu\nu\rho}\,\partial_y\xi^\nu\,\phi^{-2} 
\left(\phi D^\rho \phi +\varphi {\cal D}^\rho \varphi \right)
\;.
\label{finalAB}
\eea
Finally, we may apply yet another modification to the transformation law of $B_{\mu}$
\bea
\delta^{\rm mod}_{\xi} B_{\mu}  &\rightarrow& \delta^{\rm mod}_{\xi} B_{\mu} 
-2\varphi\,\varepsilon_{\mu\nu\rho}\,\partial_y\xi^\nu
\left(
\phi^{-2}{\cal D}^\rho\varphi-\ft12\,\varepsilon^{\rho\sigma\tau}\,F_{\sigma\tau}
\right)
\;,
\label{finalAB2}
\eea
by a term proportional to $\frac{\partial {\cal L}}{\partial B_\rho}$ which constitutes 
a separate invariance of the Lagrangian.
The resulting expressions (\ref{finalAB}), (\ref{finalAB2}) precisely reproduce the
transformation behavior of the vector fields (\ref{varA4}), (\ref{varB4}).
We have thus shown that the $(3+3)$-dimensional generalized diffeomorphisms 
that we have defined in the previous chapter, consistently reduce to the action
of the standard $(3+1)$-dimensional diffeomorphisms, once the explicit solution
of the section constraints is evaluated. This agreement holds up to transformations
of the `equations-of motion-symmetry' type, that separately leave the $(3+1)$-dimensional 
action invariant.
We recall, that in $(3+3)$ dimensions similar contributions (\ref{varAB}) proportional to the
duality equations have appeared in the derivation of the vector field transformation law.
However, unlike (\ref{eoms}), (\ref{finalAB2}), the transformation law in $(3+3)$ dimensions in
fact has no ambiguity, with the form of (\ref{varAB}) uniquely determined by gauge invariance and 
compatibility
of the transformation with the constraints (\ref{fullconstraints}) on the vector fields.

\section{Conclusions and Outlook} 
In this paper we have presented a duality-covariantization of $D=4$
Einstein gravity that is manifestly covariant with respect to the Ehlers group $SL(2,\mathbb{R})$.
To this end we performed a Kaluza-Klein inspired $3+1$ split of fields and coordinates 
in the Einstein-Hilbert action (without any truncation or assumption on the topology of 
spacetime)  and then enhanced the `internal' coordinate to $Y^M$ in the ${\bf 3}$
of $SL(2,\mathbb{R})$. The theory is subject to a number of $SL(2,\mathbb{R})$ 
covariant `section constraints', which implies that only one coordinate among the $Y^M$ 
is physical, but also that among the components of the $SL(2,\mathbb{R})$ gauge field 
$B_{\mu}{}^{M}$ only one survives. Solving the constraints accordingly and eliminating 
auxiliary fields, we recover 
$D=4$ Einstein gravity. We may also reduce to $D=3$, directly starting from our formulation, 
by setting $\partial_M=0$, after which we recover the usual $SL(2,\mathbb{R})$ invariant action. 
In this sense, our formulation explains the emergence of the hidden symmetry group 
found by Ehlers in general relativity (with one isometry) more than 50 years ago \cite{Ehlers:1957}.

As mentioned in the introduction, the truncation assumed in previous papers 
in our language amounts to keeping only the potential term,
i.e.\ the last term in (\ref{visionaryAction}).
The explicit comparison with these approaches is less straightforward, as our construction
relies on the proper normalisation of the group valued matrix ${\cal M}_{MN}$.
In particular, the specific actions of \cite{Berman:2010is,Berman:2011pe,Berman:2011cg}
carry terms that are zero when $\det{\cal M}=1$ and cannot show up in our construction.

The approach introduced here should be straightforwardly extendable to higher-dimensional 
gravity theories, in particular to 11-dimensional supergravity, in which case $SL(2,\mathbb{R})$
is enhanced to $E_{8(8)}$. Previous papers have found problems in the formulation of 
$E_{8(8)}$ covariant structures and ascribed these obstacles to the dual graviton problem. 
In contrast, our construction naturally incorporates all dual fields and we are confident  that 
it may be extended to the full 11-dimensional supergravity and yield an $E_{8(8)}$ covariant
formulation of the type (\ref{visionaryAction}).

There are various possible directions of extending the present theory. 
One problem, as in DFT, is the question whether there is any way to relax the section constraints. 
Although there is a growing body of work fearlessly
going ahead and abandoning the constraints, we believe that a proper understanding 
of how to do this consistently (that is, in a gauge invariant manner) is lacking. 
A related issue in our present theory is
that we need to impose additional (yet covariant) `section constraints'  involving the 
field $B_{\mu}{}^{M}$. This is perhaps the least satisfactory feature of our formulation, 
and one may hope that eventually it can be relaxed so that, e.g., the conditions on $B_{\mu}{}^M$
are recovered as on-shell equations. E.g.\ the first-order duality equations (\ref{duality})
obtained as field equations by variation of the Lagrangian w.r.t.~$A_\mu{}^M$ imply that only
one component of the field strength associated with the gauge field $\tilde{B}_{\mu\,M}$ is
actually non-vanishing and thus is compatible with the constraints (\ref{fullconstraints}).
It is tempting to contemplate the idea that this field equation is not only compatible with
but may in fact imply (part of) the constraints~(\ref{fullconstraints}).

Another feature that is different from DFT is that the invariance under $(2+1)$-dimensional 
diffeomorphisms parametrized by $\xi^{\mu}(x,Y)$ is highly non-manifest 
and can only be checked by a quite tedious computation. It would be desirable to 
have a formulation that makes also this symmetry manifest. In this regard comparison 
with DFT is quite illuminating in that we may also here perform a Kaluza-Klein-like 
$D=n+d$ decomposition, where $D$ is the total number of spacetime dimensions, 
and we showed in an accompanying paper that the resulting formulation 
looks very similar to the one presented here, carrying $O(d,d)$ instead of $SL(2,\mathbb{R})$
covariance  \cite{Hohm:2013new1}. In the case of DFT we can recover the fully covariant theory by simply 
reverse engineering and enhancing the group as 
 \be
  O(n-1,1)\times O(d,d)\quad \rightarrow \quad O(n+d,n+d) \ = \ O(D,D) \;, 
 \ee
doubling also the non-compact coordinates, 
thus realizing the $O(d,d)$ invariant theory as a reduced and Lorentz gauge fixed form of a fully covariant theory
with $O(D,D)$ symmetry.    
The analogous step in the $SL(2,\mathbb{R})$ invariant theory would be 
to  introduce an enlarged vielbein, say, $(3+3)$-dimensional,  
 \be
  E_{\hat{M}}{}^{\hat{A}} \ = \ \begin{pmatrix}    e_{\mu}{}^{a} & A_{\mu}{}^{M}{\cal V}_{M}{}^{A}  \\[0.5ex]
  0 & {\cal V}_{M}{}^{A}   \end{pmatrix}\;. 
 \ee
Such an ansatz indeed has the potential to generate, e.g., the ${\cal M}$-dependent term 
in $\delta_{\xi} A_{\mu}{}^{M}$, see (\ref{deltaxiA}), through compensating Lorentz gauge transformations, 
exactly as happens in DFT \cite{Hohm:2013new1}. However, it is evident that the story cannot quite be as
simple, for there is no room for the extra field $B_{\mu}{}^{M}$ and it is also not clear what kind of 
generalized diffeomorphisms should be postulated in the $(3+3)$ (or higher-)dimensional theory. 
The structure of $O(D,D)$ seems to be rather special, and it appears as if in the case of 
U-duality groups one cannot enhance the symmetry as simply. In fact, the case of 
$E_{8(8)}$ makes it evident that there is no simple (finite-dimensional) group that could incorporate 
all fields. One may be inclined to resort to one of the proposals such as 
$E_{11}$ or $E_{10}$ \cite{West:2001as,Damour:2002cu}, but then of course one would have 
to explain the fate of the infinite number of extra fields. 
 
Another improved formulation or extension of our theory might be obtained 
starting from the observation in \cite{Hohm:2005sc,Hohm:2006ud} that in the 
$(3+1)$-dimensional theory the gravity 
fields $e$ and $\omega$ and the gauge fields $A$ and $B$ fit, remarkably, into a 
Chern-Simons theory for an enhanced gauge group. While it has been known for quite 
a while that pure gravity in $2+1$ dimensions can be written as a Chern-Simons gauge 
theory, based on either the (anti-)de~Sitter or the Poincar\'e group \cite{Achucarro:1987vz,Witten:1988hc}, 
the results of  \cite{Hohm:2005sc,Hohm:2006ud} showed that this group can be extended 
by generators $Q$ and $E$, so that all the gauge fields fit into an enlarged gauge connection   
 \be
  {\cal A}_{\mu} \ = \ e_{\mu}{}^{a}P_a+\omega_{\mu}{}^{a}J_{a}+A_{\mu}\; Q+B_{\mu}\; E \;. 
 \ee
The Poincar\'e algebra of translation generators $P$ and Lorentz generators $J$ is then 
extended to a semi-direct-like product with $(Q,E)$ such that the Poincar\'e  subalgebra 
receives a non-central extension by $E$. Schematically, 
 \be\label{newalgebra}
  \big[ \, J\,,\; P\;\big] \;\sim \; P\,+\, E\;.
 \ee 
Thanks to this non-central extension there is now an invariant inner product, containing 
the pairing $\langle Q\,,E\rangle\sim 1$,  
that can be used to define a Chern-Simons action for the full algebra. 
This action precisely reproduces not only the (covariantized) Einstein-Hilbert term but also the needed 
$B\wedge F$ term. So if this construction could be extended to the $SL(2,\mathbb{R})$ 
covariant fields the full action could be written as 
  \be\label{CSspeculation}
    S \ = \ \int d^3x \, d^3Y \Big( \varepsilon^{\mu\nu\rho}\Big\langle{\cal A}_{\mu},\partial_{\nu}{\cal A}_{\rho}
    +\tfrac{1}{3}\big[{\cal A}_{\nu},{\cal A}_{\rho}\big]\Big\rangle +\tfrac{1}{16}eg^{\mu\nu}
  {\cal D}_{\mu}{\cal M}^{MN}{\cal D}_{\nu}{\cal M}_{MN} -e\,V(e,{\cal M}) \Big)\;. 
 \ee 
In this form, $D=4$ gravity would take the form of a true Chern-Simons-matter theory.  
It is clear, however, that it is not quite as simple to make complete sense of the form (\ref{CSspeculation}).
For instance, the `Lie algebra' part corresponding to $A$ is given by the C-bracket, 
which does not define an actual Lie algebra, thus requiring a suitable 
extension of Chern-Simons theory. Moreover, $B$ satisfies constraints that the other fields 
do not need to satisfy and therefore these constraint first would need to be made more democratic
among the fields. Finally, we had to replace the Einstein-Hilbert term by an improved version 
in order to keep local Lorentz symmetry, and it is not obvious how to incorporate this  
into a Chern-Simons formulation. Despite these obstacles one feels that the existence of 
an algebraic structure such as~(\ref{newalgebra}) 
cannot be a mere coincidence and should be a glimpse of some deeper structure.

Finally, let us mention that recently it has been shown that DFT can be generalized 
so that it also encodes higher-derivative $\alpha'$ corrections  \cite{Hohm:2013jaa}. 
Remarkably, in the context of such an $\alpha'$-geometry the theory is almost uniquely 
determined by its gauge structure, thus giving a new approach to determine 
the higher-derivative corrections. 
It is reasonable to expect that a similar extension exists for theories of the type 
discussed here, in particular for an $E_{8(8)}$ covariant form of 11-dimensional 
supergravity. If so this would allow us to compute the higher-derivative M-theory corrections 
in a manifestly $E_{8(8)}$ covariant fashion.

\section*{Acknowledgments}
This work is supported by the
DFG Transregional Collaborative Research Centre TRR 33
and the DFG cluster of excellence "Origin and Structure of the Universe".
We would like to thank each others home institutions for hospitality.




\providecommand{\href}[2]{#2}\begingroup\raggedright\endgroup

\end{document}